# Remodeled Relativity Theory[*]

**Abhijit Biswas, Krishnan RS Mani** [**]

Department of Physics, Godopy Center for Scientific Research, Calcutta 700 008, India

## Abstract

In this paper is presented a remodeled form of Einstein's relativity theories, which retains and incorporates only experimentally proven principles. This remodeled relativity theory follows the methodology of nature, and is based on a generalized law for spinning and rotational motions, which is in fact the conservation law of momentum vector direction, and which can be successfully used for the precision computation of planetary and lunar orbits.

The most fundamental principles of the remodeled relativity theory are the conservation laws of energy and momentum. Based on eight decades of relativity experiments, we adopted two fundamental principles: one, that energy level is the underlying cause for relativistic effects and two, that mass is expressed by the relativistic energy equation from Einstein.

From the space age ephemeris generation experience and while following nature's way to conserve energy and momentum, we found sufficient reason to replace the concept of "relativity of all frames" with that of "nature's preferred frame", which helped us to escape the dilemma, faced by Einstein till 1912, when he concluded that 'there is no way of escape from the consequence of non-Euclidean geometry, if all frames are permissible'.

Einstein formulated the general theory as a law for all coordinate systems, but physicists and astronomers have continued to adopt one or other 'ad-hoc' approach that are not in complete conformity with the basic tenets of Einstein's theories.

Based on the few well-proven basic principles mentioned above, a comprehensive remodeling effort by us led to the proposed theory that uses Euclidean space to consistently and successfully simulate numerically the results of all the "well-established" tests of the general theory at their current accuracy levels (as presented in our earlier papers), and for the precise calculation of relativistic effects observed in case of the Global Positioning System applications, the accurate macroscopic clock experiments and other tests of the special theory.

Utilizing the essential essence of Einstein's theory, and enriched by the benefits of numerous space age experiments, this theory can by its capability and consistency, be proved to avoid the inadequacies of the former.

## Abstrait.

Ce papier présente une forme remodelée de théories de relativité de Einstein, qui retient et incorpore seulement ces principes prouvés expérimentalement suivant les pas de nature, et cela est basé une loi généralisée pour les mouvements de tourner et rotation, cela est vraiment la loi de conservation de direction de vecteur d'élan, et cela pourrait être utilisé avec succès pour le calcul de précision d'orbites planétaires et lunaires.

Les principes les plus fondamentaux de la théorie de relativité remodelée sont les lois de conservation d'énergie et d'élan. Basé sur huit décennies d'expériences de relativité, nous avons adopté deux principes fondamentaux: le niveau d'énergie est la cause fondamentale pour les effets relativistes et la masse est exprimée par l'équation d'énergie relativiste de Einstein.

De l'expérience de génération de l'éphéméride d'ère spatiale et en suivant la façon de la nature pour préserver de l'énergie et l'élan, nous pourrions trouver la raison pour remplacer le concept de « la relativité de tous cadres » avec cela de « le cadre de la nature préféré », qui nous a aidé échapper le dilemme, fait face à par Einstein jusqu'à 1912, quand il a conclu qu'il y a ‹ aucune façon d'évasion de la conséquence de géométrie non-euclidien, si tous cadres sont acceptables ›.

Einstein a formulé la théorie générale comme une loi pour tous les système de coordonnée, mais les physiciens et les astronomes ont continué à adopter une ou autre approche « improvisé » qui ne sont pas en conformité complète avec les principes fondamentaux de théories de Einstein.

Basé sur de peu principes fondamentaux bien-prouvés, un effort à remodeler complètement à mené nous à cette théorie avec espace Euclidien qui peut être utilisé régulièrement avec succès pour la simulation numérique des résultats de tous les tests bien-établis de la théorie générale à leurs niveaux de précision actuels (comme présenté dans nos papiers précédents), et pour le calcul précis d'effets relativistes observés en cas des applications de Système Disposant Globales, L'expérience de l'horloge macroscopique et des autres tests de la théorie spéciale.

Consister en l'essence essentielle de la théorie de Einstein, et enrichi avec les avantages experimantals d'ère spatiale, cette théorie peut prouver sa capacité et son cohérence pour éviter les insuffisances de l'ancien.

*PACS:* 95.30.Sf; 06.30.Ft; 04.90.+e, 14.60.Ef**,** 06.30.Gv, 03.30.+p

***Key words:*** relativity theory, conservation law, generalized law of rotational motions, nature's preferred frame, Global Positioning System

## 1. Introduction

Einstein's relativity theories were highly influenced by Galileo's law of inertia and principle of relativity, although both related to only rectilinear motion. Special relativity theory (SRT) had phenomenal experimental success during the decade Einstein worked on formulation of the general relativity theory (GRT), as well as during the later decades, although some of these experiments have been interpreted in ways that do not fully conform to the basic tenets of SRT, as will be shown here. Comparatively, GRT has had very few "well-established" tests to confirm it; whereas, its essential aspects have not yet been verified from controlled experiments. The just completed Gravity Probe – B experiment tests one such aspect, namely frame-dragging, which relates to spin and rotation, but its results are awaited. Nobel laureate Professor C. N. Yang had stated that he would not be surprised at all if this experiment gives a result in disagreement with Einstein's theory .

In section 2 of this paper we present the concept of relativistic time from the point of view of the Remodeled Relativity Theory (RRT) presented in this paper, which explains consistently the relevant experimental results using the relativistic nature of time without adopting any 'ad-hoc' approach.

Another notion of Einstein's theories of relativity is the direction-dependent nature of 'Lorentz or length contraction'. No direct experimental test has confirmed the phenomenon of length contraction and its direction-dependent nature over almost a century. In section 3 of this paper we present the concept of relativistic length contraction from the point of view of RRT.

Since SRT is simpler and has been more numerously verified experimentally than GRT, another 'ad-hoc' trend adopted by a sizable section of physicists is to use SRT in place of GRT contrary to Einstein's statement that the problem of formulating physical laws for every coordinate system was solved by the GRT. All such 'ad-hoc' approaches and other inadequacies of GRT indicate the strong need for a comprehensive remodeling of the relativity theory. In the space age, the authors could find reasons for giving up the concept of "relativity of all frames" and for adopting the concept of "nature's preferred frame", while formulating the RRT. These have been presented in section 4 of this paper, along with discussions on the philosophical principles of GRT from the point of view of RRT.

## 2. Relativistic Time Dilation

The abandonment of the concept of universal time and the adoption of the concept of relativistic time in its place is so counter-intuitive that Bondi [1] was prompted to remark that relativity was at first considered shocking, anti-establishment and highly mysterious, and that all presentations intended for the general public emphasized its shocking and mysterious aspects, hardly conducive to easy teaching and good understanding.
The concept of relativistic time is being interpreted below in the perspective of an alternative relativistic gravitational model, the RRT, in a bid to develop a clearer understanding of relativistic time by eliminating its mysterious, counter-intuitive nature.

## 2.1. Historical Background

Historically, the theory of Special Relativity was one of the few highly mysterious and counter-intuitive physical theories with the 'twin paradox' (or 'clock paradox') engendered by it being the most famous and the most controversial. The paradox centers on the contention that, in relativity, either twin could regard the other as a traveler, in which case each should find the other younger — a logical contradiction. This contention assumes that the twins' situations are symmetrical and interchangeable, and this is linked to the well-known principle of "relativity of all frames" or "no preferred frame", which is a basic tenet of both SRT and GRT.
The decade after 1905 saw a number of experiments confirming SRT but which did not develop that deeper understanding of relativistic time that could have removed its mysterious and counter-intuitive aspects. Even after development of GRT by Einstein, it took more than half a century for the unprecedented stability and accuracy of atomic clocks to be achieved, and the direct experimental proof of the clock paradox using macroscopic clocks was available only in the 1970's by the Hafele-Keating [2, 3] and GP-A [4] experiments. That a section of physicists was unsure about the real nature of relativistic time even during the GPS commissioning in 1977, is clear from the remark made by Ashby [5] that when the first Cesium atomic clock was placed in orbit, it was recognized that orbiting clocks would require a relativistic correction, but there was uncertainty as to its magnitude as well as its sign, and there were some who even doubted that relativistic effects were truths that would need to be incorporated. The GPS application may be said to have led to a sweeping change in the comprehension of relativistic time, as unlike all other previous experiments the GPS application in its true sense may be considered a continuous or ongoing experiment over a wider region of space.

## 2.2. Calculation of Relativistic Time Dilations

In the RRT, the two types of relativistic time dilations, namely, the gravitational time dilation and the kinematical time dilation are calculated as follows.

### 2.2.1. Coordinate System

The right coordinate system (c.s.) is not a matter of selection but a matter of detection or identification of the natural c.s. that is related to the motion of any body in question. The natural c.s. appearing as a 'preferred frame', can rather be described as Nature's preferred frame. For example, the geocentric c.s., or to be more specific the earth centered space fixed (ECSF) frame, is the natural c.s. for moving bodies such as the flying clock in the airplane, an orbiting space vehicle of the GPS, or a natural satellite of the earth like the Moon. It may be mentioned that the "natural c.s." extends up to the limit of the "sphere of influence" [6] based on the experience and the high degree of confidence attained from interplanetary transfer operations during the space-age. Thus, the motion of any body in an apparently empty space, occurs within the gravitational field of a central body, within whose sphere of influence the first body moves or orbits. The energy level of the first body and its motion at any instant, be it a celestial body or clock, is influenced by the gravitational potential of the central body strictly according to the conservation laws of energy and, the conservation laws of linear and angular momentum. Nature executes these conservation principles in its preferred frame, which is the space fixed c.s. of the relevant gravitating central body within whose sphere of influence the first body moves or orbits.

*2.2.2. Transformation Factor for a Photon*

In an earlier paper [7], we have presented a mathematical model for numerical simulation of a photon's trajectory under gravitational influence of the sun, and the computation results for the Shapiro time delay and light deflection experiments. Equation (1) from the 'photon model' given in that paper has been reproduced below in a rearranged form, which is the energy balance equation for a photon at a radial coordinate r in the relevant natural c.s., as it travels from an infinite distance to its new location.
It may be noted here that eqn. (1) given below is a simplified version for the gravitational field of the earth ignoring the perturbing effects of the sun and the Moon (the rigorous way involves simulation of its journey from infinity through the solar gravitational field and then into the gravitational field of the earth) such that the approximation involved does not lead to errors in the significant digits for the calculation of the results of the experimental tests that will be discussed in this paper.

$$h\nu = m_r \cdot c_r^2 - \frac{G.M_E \, m_r}{r} = m_r \cdot c_r^2 \left[ 1 - \frac{G.M_E}{c_r^2 \cdot r} \right], \qquad (1)$$

where
- h = Planck's constant,
- $\nu$ = frequency of the photon at an infinite distance from any gravitating body (where the gravitational influence of any other body is also insignificant),
- $c_r$ = magnitude of photon's velocity at a distance r from the center of the ECSF frame,
- $m_r$ = relativistic mass of the photon,
- G = gravitational constant, and
- $M_E$ = mass of the earth.

The gravitational Red-shift factor appears naturally in the above energy balance equation, caused by the shift in gravitational potential energy level. Thus, the term in parentheses in the above equation is the relativistic transformation factor for a photon's energy (mass-energy). The starting assumption for RRT model that the relativistic transformation factor for time is the same as the relativistic transformation factor for energy level, has been proved by the successful computation results, which compare well with the recent experimental results for the Shapiro time delay and light deflection experiments [7]. Thus, this principle of equality of the relativistic transformation factors for energy level and time (and length), was adopted as a fundamental principle of RRT, and accordingly, the relativistic transformation factor for time for a photon moving under the gravitational influence of a central body, in the relevant natural c.s., may be written as**:**

$$dt = d\tau \, [1 - F], \qquad (2)$$

where

dt　= coordinate time, that is, the proper time at a point situated at an infinite distance from the central body, where the gravitational influence of any other body is also insignificant,

dτ　= proper time at a radial coordinate r, in the relevant natural c.s., and

F　= gravitational red-shift factor (strictly speaking, it should be called a 'blue-shift' when the photon is moving closer to the gravitating mass), given by

$$F = \left[ \frac{G.M_E}{c_r^2 .r} \right], \tag{3}$$

### 2.2.3. Transformation Factor for a Material Particle or Body, or a Moving Clock

In an earlier paper [8], we have presented a mathematical model for the numerical simulation of a planet's orbit under the gravitational influence of a central body in the relevant natural c.s., along with the computation results for the anomalous precession of the perihelion of Mercury's orbit. Equation (1) from that 'matter model' has been reproduced below to explain the case of a typical geocentric orbit in a rearranged and expanded (higher order terms are not shown here, and either these terms or the square root term itself, should be incorporated when necessary) form which is the equation for the rest mass $m_0$ of a test body at an infinite distance (where the gravitational influence of any other body is also insignificant), in terms of its relativistic mass $m_r$ as it is pulled and moved to its new location under the gravitational potential of the central body.

$$m_0 = m_r . \sqrt{1 - \beta_r^2} \approx m_r . \left(1 - \frac{\beta_r^2}{2}\right) \approx m_r . \left(1 - \frac{v^2}{2.c_r^2}\right), \tag{4}$$

where

$m_r$ = relativistic mass of the moving or orbiting body at a radial coordinate r in the natural c.s., i.e., ECSF frame,

$\beta_r$ = velocity ratio of the moving or orbiting body at a radial coordinate r, given by

$$\beta_r = \left[ v / c_r \right],$$

v　= velocity of the moving or orbiting body at a radial coordinate r, in the same natural c.s., and

$c_r$ = magnitude of light velocity at a distance r from the center of the same natural c.s.

Simulation of the photon's orbit in the same natural c.s., using the photon model mentioned in the previous subsection, reveals that the following relationship exists for the velocities of light, at the current accuracy level**:**

$$c = c_r . \left[ 1 - F \right], \tag{5}$$

where

c　= magnitude of light velocity at infinite distance (where the gravitational influence of any other body is also insignificant) from the center of the same natural c.s.

Combining equations (4) and (5), and substituting from equation (3) for F, the gravitational Red-shift factor, leads to

$$m_0 \cdot c^2 = m_r \cdot c_r^2 \cdot \left(1 - \frac{G.M_E}{c_r^2 \cdot r}\right)^2 \cdot \left(1 - \frac{v^2}{2 \cdot c_r^2}\right)$$

Rearranging, simplifying and dropping the negligible terms,

$$\frac{m_0 \cdot c^2}{\left(m_r \cdot c_r^2 - \frac{G.M_E \, m_r}{r}\right)} = \left[1 - \frac{G.M_E}{c_r^2 \cdot r} - \frac{v^2}{2 \cdot c_r^2}\right], \qquad (6)$$

Equation (6) above really represents the energy balance equation for the test body moving or orbiting at a radial coordinate r in the same natural c.s. The denominator in its L.H.S. term represents total mass-energy of the 'orbiting system' (that is, of the earth along with the particular orbiting body); the first term in the denominator represents total mass-energy of the orbiting body; whereas, the second term in the denominator represents potential energy of the orbiting system. The fact that the potential energy of the orbiting system is negative indicates that this is the case of a closed orbit, that is, this particular orbiting body is always (till an equal amount of positive energy is supplied to release this body out of the sphere of influence of the central body) bound to the central body.

It may be noted here that the simplification mentioned in the second paragraph of the previous subsection has also been followed here, and a similar rigorous method should be used as and when necessary.

It may be noteworthy to mention here that the energy balance equation represented by equation (6) above for the matter model could be arrived at just by substituting the relationship for the velocities of light from the photon model in the equation for the relativistic mass in terms of the rest mass of the test body in the matter model, thus showing the consistency of the RRT model, while indicating a good compatibility of the photon and the matter models.

The L.H.S. term of equation (6) above is the ratio of the total energy (that is, mass-energy) of the orbiting system at an infinite distance from a gravitating body, to the same at a radial coordinate r, in the natural c.s. The R.H.S. term of equation (6) thus represents the energy (that is, mass-energy) transformation factor. Thus, the transformation factor for time for a material particle or a body or a moving clock, is given by

$$dt = d\tau \cdot \left[1 - \frac{G.M_E}{c_r^2 \cdot r} - \frac{v^2}{2 \cdot c_r^2}\right], \qquad (7)$$

The second term in the parentheses in equation (7) represents the gravitational dilation term, and the third term in the parentheses represents the kinematical dilation term.

*2.2.4. Transformation Factor for Kinematical Time Dilation of a Material Particle or a Clock*

When a material particle or a clock is moving in a region where the gravitational field is very weak or insignificant, then neglecting gravitational time dilation, the kinematical time dilation can be calculated using the following equation**:**

$$dt = d\tau \cdot \sqrt{1 - \left(v^2 / c_r^2\right)}, \tag{8}$$

The equality of the relativistic transformation factor with the ratio of total mass-energy of an accelerated particle to its rest mass-energy, is well known in SRT. RRT also verifies this principle of equality of the relativistic transformation factors for energy level and time, from typical experimental data from particle accelerators, viz., the data available from CERN muon storage ring [9], for muons of momentum 3.094 GeV corresponding to the relativistic factor

$$\gamma = \left[1 \Big/ \sqrt{1 - \left(v^2 / c_r^2\right)}\right] \approx 29.3 \text{ for } \beta_r = \lfloor v / c_r \rfloor = 0.9994;$$

whereas, the ratio of total mass-energy to rest mass-energy of muons, equals 29.3.

**2.3. Review of some notable Relativistic Time Dilation Experiments**

Among all the paradoxes engendered by the relativity theory, the most famous and controversial is the "twin paradox". French [10] critically remarked in his 1966 book on SRT that it has been argued by some writers that an explanation of the twin paradox must involve the use of GRT on the ground that the moving clock had phases of acceleration and deceleration. He further stated that SRT can do the job of predicting the time lost due to kinematical time dilation without bringing in the notions of equivalent gravitational fields. The fact that accelerations do not affect the clock rates became known at the beginning of the nineteen sixties, from experiments utilizing the M**ö**ssbauer effect, where the temperature-dependent (e.g., Pound and Rebka) experiments [11] demonstrate that accelerations of the order of $10^{16}$g, arising from lattice vibrations, produce no intrinsic frequency shift in $Fe^{57}$ nuclei to an accuracy of the order of 1 part in $10^{13}$.

Resnick [12] compared the cases of different aging of twins due to slowing down of human life processes by refrigeration with the case of the different aging of twins due to relativistic motion, and explained that the different aging in the twin paradox was due to the difference in motion but the paradox in the relativistic case was that the situation appeared to be symmetrical (incorrectly) because (uniform) motion was relative. He added that just as the temperature differences were real, measurable, and agreed upon by the twins in the foregoing example, so were the differences in motion real, measurable, and agreed upon in the relativistic case — but the changing of inertial frames, that is the accelerations, were not symmetrical. In continuation Resnick [12] has further stated that the results were absolutely agreed upon, and has brought in the typical GRT-based argument involving the accelerating-decelerating phases of the moving spaceships to be the cause for the time dilation, even though acceleration had been proven to cause no time dilation [13, 14].

In RRT, while accepting that "the differences in motion are real and measurable in the relativistic case", we do not concur in the opinion that such results can be explained while sticking to the principle of 'relativity of all c.s.' This will be shown in the subsections below that explain the moving clock experiments, and the time adjustment on a continuous basis of GPS clocks.

Einstein [15] explained his view on the concept of twin paradox and its relationship with the principle of 'relativity of all c.s.', using a thought experiment; he pointed out that an observer at rest in the lower c.s. would find that a moving clock changed its rhythm. Contrary to the modern views mentioned above, he [15] specially remarked that the same result could certainly be found if the clock moved relative to an observer at rest in the upper c.s., and that the laws of nature must be the same in both c.s. moving relative to each other.

Thus, Einstein can be said to have not only stated unambiguously that the situation of the two clocks (or, the twins) to be symmetrical, but also reinforced this statement by linking it with his second postulate of SRT. He thus refuted the contemporary view on the twin paradox that the situation of the two clocks (or, the twins) is not symmetrical. But, contrary to his view and SRT, the macroscopic clock experiments (including GPS) as explained below, have proven that the twin's situation is not symmetrical. In other words, the macroscopic clock experiments have clearly contradicted Einstein's SRT view and supported the RRT view.

Einstein stated [15] that only relative motion could be observed, as one could not talk about absolute uniform motion because of the Galilean relativity principle. He further mentioned [15] that as motion is relative and any frame of reference can be used, there seems to be no reason for favoring one c.s. rather than the other.

But in the space-age, RRT has found this reason (as explained at subsections 2.4.2 and 4.2) and adopted the nature's preferred-frame. The explanations given in the subsections and sections below show how experimental results on time dilation support the RRT concept.

*2.3.1.   Measurements of the lifetime of Cosmic-ray Muons*

In 1941 B. Rossi and D.B. Hall conducted a classic time dilation experiment using the cosmic ray muons, whose modern filmed version [16] was conducted in 1963.
The cosmic ray muon experiment is a successful demonstration of both SRT and RRT as regards the equation for kinematical time dilation, as given by

$$dt = d\tau \cdot \sqrt{1 - \left(v^2 / c_r^2\right)}, \qquad (9)$$

According to RRT, we interpret the experimental data as follows**:**
For a detailed and precise calculation of muons falling from an altitude under the influence of the gravitational field of earth, it is necessary to consider their motion in the ECSF frame. However, one can simply calculate only kinematical time dilation for this case, since the effect of gravitational time dilation is insignificant.

*2.3.2.   Round trip experiments using Muons*

In this experiment, the relativistic time dilation for positive and negative muons in a circular orbit were measured in the CERN storage ring. At a muon speed corresponding to a relativistic factor $\gamma = 29.33$, the measured lifetime of the positive muons was found to be in accordance with SRT and the measured lifetime at rest; the relativistic time dilation factor agreed with experiment with a fractional error of 1.6E-3 to 2.0E-3 at 95% confidence level. Even at accelerations as large as ~$10^{18}$ g, no effect was observed on particle lifetime. [9]

Thus, for accelerations as large as ~$10^{18}$ g, this experiment confirms the clock hypothesis, which states that the tick rate of a clock when measured in an inertial frame depends only upon its velocity relative to that frame, and is independent of its acceleration or higher derivatives.

This experiment involving a round trip by the muons at a relativistic speed of (v ~ 0.9994 c), is one of the few experiments that simulate closely the so-called 'twin paradox' which was discussed in Einstein's first paper [17].

For this case, the RRT, like the SRT, takes into account only the kinematical time dilation for muons for explaining the experimental results.

### 2.3.3. Experiments using the Mössbauer effect

Two experiments using the Mössbauer effect are being briefly discussed here, one using thermal energy for creating the frequency shift while the other uses the mechanical energy.

In the first experiment, Pound and Rebka [13] utilizing the recoil-free emission of gamma rays, namely the Mössbauer effect, in $Fe^{57}$ and taking advantage of the difference in nuclear motion associated with the temperature difference between the source and the absorber, concluded that their experiment demonstrated the second order Doppler effect. Sherwin while discussing this experiment in a review paper [11] concluded that the observed fractional frequency shift, which occurs between the source and the absorber, is in quantitative agreement with the generally accepted calculations for the 'clock paradox'.

In the second experiment, Hay et al [14] utilizing the Mössbauer effect in $Fe^{57}$ and employing mechanical rotation to produce the necessary acceleration, mentioned that the fractional frequency shift could be calculated either by treating the acceleration as an effective gravitational field and calculating the difference in potential between the source and the absorber, or by using the time dilation of SRT.

In the Pound-Rebka [13] experiment, the observed frequency shift which for the former is a consequence of the kinematical time dilation caused by the motion of the nuclei in the crystal lattice due to thermal vibrations; whereas, in the Hay et al [14] experiment, the observed frequency shift is a consequence of the kinematical time dilation caused by the energy shift due to the mechanical motion of the absorber relative to the source.

The common point of both the experiments mentioned above, is that 'the fractional frequency shift is equal to the fractional energy shift'. This principle corresponds to a basic principle of RRT, according to which the relativistic effects are caused by variations in energy level.

### 2.3.4. Around-the-world atomic clock flights by Hafele-Keating

Hafele-Keating conducted the first experiment to measure the relativistic time delay effects on macroscopic clocks that were flown on commercial jet flights around the world twice, once eastward and once westward.

To evaluate the relative timekeeping behavior of flying clocks and ground reference clocks at the USNO (U.S. Naval Observatory) by reference to hypothetical reference clocks of an inertial (non-rotating) reference space [2], Hafele-Keating chose the ECSF frame as an inertial frame of reference. Noting that flying clocks lost time during the eastward trip, and gained time during the westward trip, they concluded that effects of travel on the timekeeping behavior of macroscopic clocks are in reasonable accord with predictions of the conventional theory of relativity [3]

Hafele-Keating has mentioned while explaining the theory [2], that SRT predicts that a moving standard clock will record less time compared with coordinate clocks distributed at rest in an inertial reference space. While they have not mentioned their reason for choosing the ECSF frame as an inertial frame of reference instead of any other inertial frame of reference, viz., 'sun centered space fixed' (SCSF) frame or a solar 'bary-centric space fixed' (BCSF) frame, it is obvious that the reason was that their result could be fitted only in the ECSF frame, which according to the RRT model is the Nature's preferred frame for this case.
Their experiment while not specifying any norm for the choice of a preferred frame has at the same time rejected Einstein's basic premise of "relativity of all frames".

*2.3.5.   Gravity Probe A*

Vessot et al [4] conducted the Gravity Probe A (GP-A) experiment, which is also known as the "Rocket Red-shift experiment" or the "Clock Experiment", in 1976, using hydrogen maser clocks. The observed data was compared with the frequency variations predicted by an equation derived in their earlier paper [18]. This equation corresponds to RRT equation (7) derived above at subsection 2.2.3, and is reproduced below.

$$dt = d\tau \cdot \left[ 1 - \frac{G.M_E}{c_r^2 \cdot r_s} - \frac{v_s^2}{2 \cdot c_r^2} \right], \qquad (10)$$

where
$v_s$ = velocity of the rocket clock in ECSF frame, and
$r_s$ = radial coordinate of the rocket clock in ECSF frame.

In the equation above, the second and the third terms in the parentheses correspond to the gravitational and the kinematical time dilations respectively, of the rocket clock with respect to the coordinate clock in the ECSF frame.
Just as in the Hafele-Keating experiment, here too the experimenters had to fit their data with respect to the ECSF frame as an inertial frame of reference, which can be said to be a preferred-frame according to GRT terminology, and to be a nature's preferred-frame according to RRT.

*2.3.6.   GPS*

GPS is a state-of-the-art timekeeping system [5] with high precision stable atomic clocks, clarifying man's understanding of the concept of relativistic time much like an on-going experiment. Out of its three segments, the Space segment consists of a network of 24 satellites in nearly circular orbits, which are distributed such that from any point on the earth, four or more satellites are almost always above the local horizon. Tied to their clocks are signals that are transmitted from each satellite, which like sequences of transmission events in space-time, are characterized by positions and times of transmission and associated with messages specifying the transmission events' space-time coordinates. The Control segment comprises of a number of ground-based monitoring stations, which continually gather information from the satellites. These data are sent to a Master Control Station, which analyzes the constellation and projects the satellite ephemerides and clock behavior forward for the next few hours. This information is then uploaded into the satellites for retransmission to users. The User Segment consists of all users or receivers near the earth, which can receive signals transmitted from the satellites, and can determine its position and time by decoding navigation messages from four satellites to find the transmission event coordinates, and then solving four simultaneous one-way signal propagation equations. Thus they are able to determine their position, velocity, and the time on their local clocks.

Ashby [19] mentions that GPS clocks are synchronized in the earth centered inertial (ECI) frame, in which self-consistency can be achieved. To both the satellite clocks, and the ground-based clocks of GPS, a set of appropriate corrections is applied, based on the known positions and motions of the clocks. Thus, GPS clocks' time is "coordinate clock time", since at each instant, it agrees with a fictitious atomic clock at rest in the ECI frame, whose position coincides with the earth-based standard clock at that instant.

The average frequency shift due to relativistic effects on satellite clocks in orbit according to Ashby [20], is corrected downward in frequency by 446.47 parts in $10^{12}$, which is a combination of five different sources of relativistic effects**:** gravitational frequency shifts of ground clocks due to earth's monopole and quadrupole moments, gravitational frequency shifts of the satellite clock, and second-order Doppler shifts from motion of satellite and earth-fixed clocks.

Using the RRT model, the sum of the above-mentioned five different sources of relativistic effects on satellite clocks has been computed using appropriately the equations given at subsection 2.2.3, in the ECSF frame, which is the natural c.s. for the satellite clocks, and the resulting average frequency shift of these clocks was found to be 446.4723 parts in $10^{12}$. It may be mentioned here that in the RRT, the speed of light has been adopted to have the standard magnitude of 299,792,458 m/sec. at mean sea level (MSL) on earth while for the GPS calculations, the magnitude of the speed of light $c_r$ at any radial coordinate 'r' can be determined precisely by numerically simulating a photon's trajectory in the ECSF frame. Its approximate value may however, be calculated from equation (5) given above. The speed of light is variable but at the current level of accuracy of GPS clocks, its effect on the average frequency shift is insignificant since it affects only the tenth significant digit.

We find that here too the GPS relativists could achieve self-consistency by synchronizing GPS clocks in the ECI frame or the ECSF frame, which is a preferred-frame according to GRT terminology and nature's preferred-frame according to RRT.

The GPS relativists, in applying the present relativity theory during GPS commissioning, faced some problems and that had to be overcome. In this connection,

Ashby mentions [19] that in an inertial frame, a network of self-consistently synchronized clocks can be established either by transmission of electromagnetic signals that propagate with the universally constant speed c (that is, Einstein synchronization), or by slow transport of portable atomic clocks; whereas, in a rotating reference frame, the Sagnac effect prevents a network of self-consistently synchronized clocks from being established by such processes. He adds that this is a significant issue in using timing signals to determine position in the GPS, as the Sagnac effect can amount to hundreds of nanoseconds; and a timing error of one nanosecond can lead to a navigational error of 30 cm.

He also mentions [19] that to account for the Sagnac effect, a hypothetical non-rotating reference frame, namely, the so-called ECI Frame is introduced, and, time in this frame is adopted as the basis for GPS time. He also remarks [19] that although the earth's mass encompasses the origin of the ECI frame and has significant gravitational effects, the gravitational effects in the GPS, to an extremely good approximation, can be simply added to other effects arising from the SRT.

From the above, it can be said that application of GRT revealed its inherent inconsistency while applying it for GPS, which is a gravitational problem in a rotating frame.

## 2.4. Discussion

### 2.4.1. Fundamental Principles of the RRT

Since the conservation laws of energy and, of linear and angular momentum are valid in both the macroscopic and microscopic realms, these have been adopted as the fundamental principles of the RRT.

Another fundamental principle of the RRT is that the relativistic effects like time dilation, length contraction, etc. are caused by a variation in the energy level (due to all forms of energy). The time-related relativistic effect is evident in two commonly observable phenomena**:**
- gravitational time dilation — It can be clearly seen from equations (1) and (6) that gravitational red shift appear in these energy balance equations naturally, and is related to variation in energy level.
- kinematical time dilation — The fact that the kinematical time dilation is caused by variation of kinetic energy level, is well known from the experience of particle accelerators and from many experiments.

The RRT model is thus based on the fact that the relativistic transformation factor for time (as well as for length) is the same as the relativistic transformation factor for energy (that is, mass-energy) level. This principle has been proved from experiments as well as by successful computation results, which compare well with the recent experimental results for the Shapiro time delay and light deflection experiments [7], as well as with the recent observational results for the anomalous precession of the perihelion of Mercury's orbit [8].

Another fundamental principle adopted in RRT, is the well-proven relativistic energy equation from Einstein, given by**:**

$$E = m_r \cdot c_r^2, \tag{11}$$

where
- $E$ = total mass-energy of a moving or orbiting body at a radial coordinate r in the natural c.s.,
- $m_r$ = relativistic mass of the same moving or orbiting body at the same location, and
- $c_r$ = magnitude of light velocity at the same location in the same natural c.s.

### 2.4.2. *Choice of Coordinate System*

In the pre Space age scenario, Einstein [15] stated that as motion is relative and any frame of reference can be used, there <u>seems to be no reason</u> for favoring one c.s. rather than the other.

Space age computational capability and accuracy in celestial mechanics has taught us that the LSA derived astronomical constants (e.g., the planetary masses, etc.) of nature, which are an outcome of global fits done during the generation of a particular ephemeris, are consequences of not only the gravitational model, but also of the coordinate frame. In other words, the constants of nature are linked to the coordinate frame, which means that today one has to accept the existence of the constants of nature as a concomitant of only one appropriate preferred frame and the relevant orbit or orbits, linked to them. This 'preferred frame' according to the RRT is necessarily the "nature's preferred frame"; whereas, SRT and GRT, born prior to the space age, started with the concept of "relativity of all frames".

The importance of "nature's preferred frame" can also be understood using another approach from the energy-balance equation (6) given above. The denominator in this equation represents total mass-energy of the 'orbiting system'. The first term in the denominator represents total mass-energy of the orbiting body; whereas, the second term in the denominator represents potential energy of the orbiting system, which includes the central body along with the particular orbiting body. This orbiting system relates to a "nature's preferred frame", in which the central body and the orbiting body are intimately linked through their energy-interaction. In other words, nature itself applies the conservation laws of energy and, of linear and angular momentum for this particular 'orbiting system' whose central body's center-of-mass is the origin of the "nature's preferred frame".

Thus, for calculations using the RRT, the first and most important step is to the identify nature's preferred frame even for problems outside celestial mechanics, that is, for all problems that involve application of the conservation laws of energy and, of linear and angular momentum.

Observer's c.s., or a c.s. chosen by an observer may be useful in some limited applications, but certainly not in case of relativistic gravitation, where one can not choose a c.s., but can only identify a c.s. that nature has already chosen.

### 2.4.3. *Speed of Light*

In the RRT, the magnitude of the speed of light $c_r$ varies with the radial distance from the center of a gravitating mass or from the origin of the natural c.s. However, the concept of variable speed of light is not a postulate of the RRT. The photon is treated as a particle similar to any matter particle moving under the gravitational influence of the sun and the planets of the solar system, the only difference being its total energy as given by Planck's quanta which is related to its relativistic mass by the famous

Einstein equation, as given at equation (11) above. In fact, our mathematical model for the photon program is so similar to the program for the material particles and celestial bodies that the RRT model can be said to relate the macroscopic and microscopic realms together.

Successful numerical simulation results using our photon and matter models [7, 8] validated the concepts that had gone into the mathematical model, and the variable speed of light $c_r$, was a consequence of this simulation.

Thus, according to the RRT, velocity of light in vacuum or empty space, is intimately linked to gravitational fields and gravitational potential energy level at any point in space, and may be determined precisely from simulation and approximately from equation (5) given above.

Einstein [21] had asserted that the law of the constancy of the velocity of light in vacuum, which constituted one of the two fundamental postulates of the SRT, was not valid according to the GRT, as light rays could curve only when the velocity of propagation of light varies with position. He concluded [21] that the SRT could not claim an unlimited domain of validity; as its results held only so long as it was possible to disregard the influences of gravitational fields on the phenomena of light propagation.

In fact, the 'variable speed of light' (VSL) is an outcome of the fundamental principles of the RRT model, and can be said to have provided a mathematical model corresponding to Einstein's above conclusion regarding the influences of gravitational fields on the speed of light, by enabling one to numerically simulate the magnitude of the speed of light, varying with position.

Thus, RRT can be said to have avoided the superfluous postulate of "constancy of light velocity" of SRT, and to have computed the light velocity using its Photon model [7], which utilizes an Einsteinian [15] concept that a beam of light will bend in a gravitational field exactly as a material body would if thrown horizontally with a velocity equal to, that of light, and thus treats the motion of photon under the influence of gravitational field at par with any particle having mass, and that does not use any additional postulate or principle, but uses only those principles or basic equations that have been used by A.H. Compton in 1923, for derivations of the Compton Effect. In the process, it proves the above-mentioned principle stated by Einstein [21] that the gravitational bending of light rays is the effect of variation with position of the velocity of propagation of light under the influence of gravitational field.

We propose an experiment that will verify the magnitude of the velocity of light at locations closer to the sun. Such an experiment can be planned from an Orbiter around or a Lander on, any of the planets**:** Mercury (preferable) or Venus, or in the solar probe mission (more preferable, if it takes off early), where the experimental result will clearly prove whether GRT or RRT is the right relativistic gravitational model.

*2.4.4. Application of GRT in GPS*

As the earth-centered earth-fixed (ECEF) frame is of primary interest for navigation and the GPS operates in the gravitational field of the earth, GRT is thus supposed to be the relativity model that is appropriate to tackle the gravitational problem in a rotating ECEF frame. Since almost all users of GPS are at fixed locations on, or are moving very slowly over, the surface of the rotating earth, an early GPS design decision was taken [5] to broadcast the satellite ephemerides in a model ECEF frame, designated by the symbol WGS-84. Ashby mentioned that the receiver must generally perform a different rotation for each measurement made, into some common inertial frame, so that the navigation equations (12) mentioned below apply; and, after solving the propagation delay equations, a final rotation must usually be performed into the ECEF to determine the receiver's position. He concluded that this can become exceedingly complicated and confusing. Ashby further mentioned [5] that it was necessary to invoke the inertial frame for GPS application, for the following reasons**:**

- o many physical processes (such as electromagnetic wave propagation) are simpler to describe in an inertial reference frame,
- o the inertial frames are certainly needed to express the navigation equations (that is, the four simultaneous one-way signal propagation equations), viz.,.

$$c^2 \cdot (t - t_j)^2 = |\mathbf{r} - \mathbf{r}_j|^2, \quad j = 1, 2, 3, 4., \quad (12)$$

as it would also lead to serious error in asserting the above equation in the rotating ECEF frame, and
- o Simple-minded use of Einstein synchronization in the rotating frame leads to a significant error, because of the additional travel time required by light to catch up to the moving reference point, as compared to the travel time required in the underlying inertial frame.

He further adds [5] that synchronization of GPS clocks is thus performed in an underlying inertial frame (ECI) in which self-consistency could be achieved.

In this connection, Ashby also mentions [19] that in a rotating reference frame, the Sagnac effect prevents a network of self-consistently synchronized clocks from being established by transmission of electromagnetic signals. Ashby further mentions [19] that as a consequence of the Sagnac effect, observers in the rotating ECEF frame using Einstein synchronization will not agree that clocks in the ECI frame are synchronized, due to the relative motion. He also remarks [19] that observers in the rotating frame cannot in fact even globally synchronize their own clocks, due to the rotation.

### 2.4.5.  *Inadequacies of GRT*

As mentioned in the previous subsections, the practical difficulties faced by GPS relativists, when using a ECEF frame are a clear indication of the limitations and inadequacy of the GRT while solving this gravitational problem in a rotating frame. While commenting on the concept of "relativity of all frames", Einstein [15] stated that he wanted to formulate physical laws that are valid for all c.s., not only for those moving uniformly, but also for those moving quite arbitrarily, relative to each other. He felt that it would be possible to apply such laws of nature to any c.s., and that the two sentences, "the sun is at rest and the earth moves," or "the sun moves and the earth is at rest", would simply mean two different conventions concerning two different c.s. He added that really relativistic physics must apply to all c.s. and, therefore, also to the special case of the inertial c.s., and hence GRT, the new general laws valid for all c.s. must, in the special case of the inertial system, reduce to the old laws known as the SRT.

It is seen that Hafele-Keating, Vessot et al, and GPS relativists have all found the non-rotating ECI frame useful and hence chose it as a preferred frame, even though they did not give their reasons for the choice of ECI instead of any other inertial reference frame. In short, all of them used SRT and GRT arguments with a preferred inertial frame instead of using GRT for their gravitational problem in the rotating ECEF frame. Ashby [19] stated that observers in the rotating frame in fact cannot even globally synchronize their own clocks, due to the rotation, as explained at subsection 2.4.4.

Even astronomers working on ephemeris generation have not been using GRT in its purest form. They find it useful to employ GRT equations in a "preferred frame" for planetary and lunar orbits.

Since, SRT is simpler and has been more numerously verified experimentally than GRT, this trend of adapting and using SRT in lieu of GRT started since the nineteen-twenties, with the relativistic electron theory of P.A.M. Dirac, relativistic quantum field theory such as QED, etc., and this trend continues to date.

We find thus that while some physicists and astronomers used only SRT instead of GRT; and some used SRT or GRT with a preferred frame, others used both SRT and GRT arguments but with a preferred frame.

In this centennial year of the SRT, the crisis in the GRT has been described in the following words in the web page of Stanford University on Gravity Probe – B [22]**:**

> "**….** it (GRT) remains one of the least tested of scientific theories. General relativity is hard to reconcile with the rest of physics, and even within its own structure has weaknesses. Einstein himself was dissatisfied, and spent many years trying to broaden his theory and unify it with just one other branch of physics, electromagnetism. Modern physicists seeking wider unification meet worse perplexities. Above all, essential areas of general relativity have never been checked experimentally."    **….**
>
> "**….** Other more profound phenomena, however, remain untested. Save for some indirect evidence from the binary pulsar, no data exist on gravitational radiation. Even less is known about a vitally important relativistic effect — "frame-dragging". Moreover, deep theoretical problems — some old and some new — remain. Einstein himself remarked that the left-hand side of his field equation (describing the curvature of space-time) was granite, but that the right-hand side (connecting space-time to matter) was sand. The mathematical structures of general relativity and quantum mechanics, the two great theoretical achievements of 20th century physics, seem utterly incompatible. Some physicists, worried by this and by our continued inability to unite the four forces of nature — gravitation, electromagnetism, and the strong and weak nuclear forces — suspect that general relativity needs amendment. **….**
>
> Grand unification is the greatest challenge confronting theoretical physicists today. Gravitation, the strong nuclear forces, and the partially unified electro-weak forces must be connected, but how? Even the issues remain speculative but several clues suggest that general relativity may require amendment  **….**"
>
> "….. Even the issues remain speculative but several clues suggest that general relativity may require amendment, and that the amendment, in the words of Nobel laureate C. N. Yang 'somehow entangles spin and rotation.' Says Yang: 'Einstein's general relativity theory, though profoundly beautiful, is likely to be amended.... That the amendment may not disturb the usual tests is easy to imagine, since the usual tests do not relate to spin [i.e. frame-dragging]. The Stanford experiment is especially interesting in that it *focuses on the spin.* I would not be surprised at all if it gives a result in disagreement with Einstein's theory.' ...."

*2.4.6. Law for Spinning and Rotational Motions*

In keeping with the general tenor of Professor Yang's views mentioned above, the RRT model uses a generalized vector law of spinning and rotational motions, that has been derived directly from the fundamental conservation laws of momentum, and that expresses a high degree of symmetry for all types of spinning and rotational motions. From the way this law has been derived at Appendix I below, it can be seen that it is in fact the conservation laws of linear and angular momentum vector direction, expressed in a generalized form.
This RRT law is valid in both the macroscopic and microscopic realms, as it is born from and hence inherits the nature of the fundamental conservation laws of momentum.
According to the GRT or the Machian concept in contrast with the RRT law of rotational motions, the inertial force is caused by the rotation of a body with respect to the totality of other ponderable bodies in the universe.
In RRT model, the LHS and the RHS parts of the equations of motion consist respectively of the gravitational and the inertial forces. The equations for inertial forces in RRT model, are obtained from the generalized law of rotational motions. The LHS part of this equation of motion consisting of the resultant gravitational forces due to the n-body effect, is written with respect to the relevant natural c.s., and is numerically integrated to obtain the epoch-wise position and velocity vector data of any orbiting planet of the solar system. The Ordinary Differential Equations (ODE's) corresponding to the RHS part of this equation of motion consisting of the inertial forces is numerically integrated and this enables us to verify the centennial precession rate of a particular orbiting planet due both to the relativistic effects and to the planetary perturbations, and also enables us to ensure that the conservation laws of energy, linear and angular momentum are obeyed during all stages of epoch-wise calculations. This procedure ensures that we are following the methodology of nature in RRT.

But no such application of the Machian concept for precision computation of planetary orbits could be found in the relevant literature.
It is the contention of the authors of this paper that these inertial forces or torques are an additional category of the fundamental interactions and these merit a new classification as the "conservation forces".
- as their equations are expressed by a generalized vector law that is in fact the conservation law of momentum vector direction,
- as they act as a mechanism of Nature for conservation of the momentum vector direction, by resisting any change of momentum vector direction,
- as they have their origin in fundamental conservation laws which hold good in both the macroscopic and the microscopic realms, and,
- as they are represented by a single generalized vector law that expresses a high degree of symmetry among all types of spinning and rotational motions.

## 3. Relativistic Length Contraction

The concept of length contraction is an important consequence of the present relativity theory.  Among the three principal relativistic effects proposed by Einstein a century ago, the time dilation and relativistic mass increase had been directly verified by many experiments. But, the phenomenon of length contraction along only the direction of motion has defied verification till date by experimentation. This paradoxical concept, which specifies length contraction only along the direction of motion with no contraction along the transverse direction, has been primarily responsible for making the Einsteinian concept of space a non-Euclidean one. In this section, we discuss this important aspect of the present relativity theory, and also present the views of the RRT on its various ramifications.

### 3.1. Historical Background

In the eighteenth and early nineteenth centuries, the three absolutes of Newton were space, time and mass. Then came Maxwell, who made the velocity of light, c, central to electromagnetism. At the beginning of the twentieth century, Einstein sensed that the tide of discoveries in electromagnetism indicated the decline of the mechanical view. Where Maxwell's and Newton's ideas came into conflict, Einstein modified the ideas of Newton to fit Maxwell's. His SRT replaced Newton's three absolutes by a single one, the velocity of light, and established two paired quantities, space-time and mass-energy, related through c. In Einstein's SRT, mass and energy are inter-convertible through the relativistic energy expression**:**      $E = mc^2$.
While formulating SRT, Einstein replaced the classical transformation law with the relativistic transformation law called the Lorentz transformation, which incorporated the concept of constancy of the velocity of light and led to relativistic transformation factors as a consequence of the relative velocity of the moving frame, or in other words, as a consequence of the translational kinetic energy of the moving frame. The SRT notion of 'length contraction' only along the direction of motion is a consequence of the Lorentz transformation, and is paradoxical in nature.

### 3.2. Mathematical Model and Terms related to Length Contraction

*3.2.1. Length Contraction*

Using matter model [8] and based on the equations (4) and (6), given in section 2 above, for a test body as it is pulled and moved to its new location under the gravitational potential of the earth, the energy balance equation may be written in the ECSF frame as follows**:**

$$\frac{E_0}{E_r} = \frac{m_0 \cdot c^2}{\left(m_r \cdot c_r^2 - \frac{G.M_E \, m_r}{r}\right)} = \left[1 - F_r\right] \cdot \sqrt{1 - \beta_r^2} \,, \tag{13}$$

where
$E_0$ = rest mass-energy of the moving or orbiting body, at infinite distance (where the gravitational influence of any other body is insignificant)
$E_r$ = mass-energy of the moving or orbiting body at a radial coordinate r in the natural c.s., i.e., ECSF frame,

$c_r$ = magnitude of light velocity at a distance r from the center of the ECSF frame, that is, at the location of the particle accelerator, and

$F_r$ = gravitational red-shift factor, given by

$$F_r = \left[ \frac{G.M_E}{c_r^2 .r} \right], \qquad (14)$$

other notations being the same as given earlier for equation (4).

For any material body or particle moving or orbiting in the ECSF frame, or even in the SCSF frame, under the gravitational attraction of the earth or the sun respectively, the order of magnitude of the terms $F_r$ and $\beta_r$ is very small. Thus, from gravitational experiments in the solar system (where the highest possible magnitude of $F_r$ is only a few parts per million), it is not possible either to directly get verification, or even to indirectly get an order-of-magnitude measure, of the phenomenon of length contraction.

However, at the earth-based accelerator laboratories, it is possible to understand the effect of length contraction, at relativistic speeds. The particles are accelerated by the electromagnets in these accelerators, which form the natural c.s. for the accelerated particles, since the variation in gravitational potential energy level is negligible. For such a case, equation (13) takes the simplified form,

$$\frac{E_0}{E_r} = \sqrt{1 - \beta_r^2}, \qquad (15)$$

where

$E_0$ = rest mass-energy of the moving particle in its own frame where it is at rest,

$E_r$ = mass-energy of the moving particle at the location of the particle accelerator in its natural c.s.,

other notations being the same as given earlier for equation (4).

*3.2.2. Resolution*

To look at the constituent parts of matter, it is necessary to break it apart in the particle accelerators, for which a quest is on for ever-higher energy accelerators. The reason for this, lies in the essential wave-particle duality and the fact that the attainable resolution of any type of imaging system is dependant on the order of the wavelength of the waves used for the imaging. Thus, the electron microscopes have higher resolution because the associated wave properties of the electrons have much shorter wavelengths (of the order of $10^5$ times shorter than visible light) than the optical microscopes. The de Broglie relationship, which allows one to calculate these wavelengths, implies higher the momentum (and thus a higher mass-energy) of the electron shorter its wavelength.

$$\lambda_r = \frac{h}{m_r \cdot v_r} = \frac{h \cdot c_r}{\beta_r \cdot E_r}, \qquad (16)$$

where

$E_r$ = mass-energy of the moving particle at the location of the particle accelerator in its natural c.s., and

$c_r$ = magnitude of light velocity at a distance r from the center of the ECSF frame, that is, at the location of the particle accelerator, other notations being the same as given for earlier equations.

The extremely small nuclear size implies that much higher energy particles need to be used to resolve details of its structure, and to probe smaller detail will require higher and higher energy particles to do the imaging. The highest energy scattering experiments to date reach a resolution of about a thousand times smaller than a proton. The resolution wavelength $\lambda_r$ of the waves of matter particle indirectly gives an order-of-magnitude measure of the diameter of the matter particle, and thus indirectly confirms the phenomenon of length contraction. The resolution wavelength $\lambda_r$ of the waves of the same matter particle, at its own rest frame, can then be expressed as:

$$\lambda_0 = \frac{h \cdot c_r}{\beta_r \cdot E_0} , \qquad (17)$$

where

$E_0$ = rest mass-energy of the moving particle in its own frame where it is at rest,

Dividing equations (16) by (17), and substituting the transformation factor for energy from equation (15) while noting that gravitational effect is negligible here, leads to

$$\frac{\lambda_r}{\lambda_0} = \frac{E_0}{E_r} = \sqrt{1 - \beta_r^2} , \qquad (18)$$

The above equation proves that the transformation factor for length contraction is the same as that for energy.

### 3.2.3. Range of Forces

In case of nuclear particles, a relativistic particle's length contraction can be understood in terms of 'range of forces', a commonly used concept of quantum mechanics. According to this concept, if a force involves the exchange of a particle, that particle has to "get back home before it is missed" in the sense that it must fit within the constraints of the uncertainty principle. A particle of relativistic mass-energy $E_r$ can be exchanged if it does not go outside the bounds of the uncertainty principle in the form of the following equation:

$$\Delta E \cdot \Delta t = E_r \cdot \Delta t \geq \frac{h}{4\pi} , \qquad (19)$$

Two limiting cases of the above equation are $E_0 \cdot \Delta t = \frac{h}{4\pi}$, and $E_r \cdot \Delta t > \frac{h}{4\pi}$, where $E_0$ represents the rest mass-energy of the particle.

A particle, which can exist only within the constraints of the uncertainty principle, is called a "virtual particle", and the time in the equation above represents the maximum lifetime of the virtual exchange particle. Since this exchange particle cannot exceed the light speed limit $c_r$ at that location, it cannot travel further than $c_r$ times that lifetime. The maximum range of the force would then be of the order of

$$R_0 = c_r \cdot \Delta t = \frac{h \cdot c_r}{4\pi \cdot E_0}, \tag{20}$$

The lower magnitude of the range of the force can then be written as

$$R_r = c_r \cdot \Delta t > \frac{h \cdot c_r}{4\pi \cdot E_r}, \tag{21}$$

Dividing equations (21) by (20), and substituting the transformation factor for energy from equation (15) while noting that gravitational effect is negligible here, leads to

$$\frac{R_r}{R_0} = \frac{E_0}{E_r} = \sqrt{1 - \beta_r^2}, \tag{22}$$

The range of a force term $R_r$ that corresponds to a "virtual particle" of matter, indirectly gives an order-of-magnitude measure of the diameter of the particle, and thus indirectly confirms the phenomenon of length contraction. Equation (22) above proves that the transformation factor for length contraction is the same as that for energy.

### 3.2.4. Compton Wavelength

A.H. Compton conducted the photon-electron scattering experiment in 1923, by allowing a beam of X-ray photons of sharply defined wavelength to fall on a graphite block, and by measuring the intensity of the scattered X-rays as a function of their wavelength, for various angles of scattering. His interpretation in exceptionally modern terms can be viewed as the earliest application of the RRT model. He considered photon as a particle of relativistic mass (equivalent to its mass-energy) interacting with an electron (a matter particle) and analyzed the results, as in RRT model, only on the basis of conservation laws of energy and momentum. Compton shift, the wavelength shift in such scattering, depends only upon the angle of scattering for a given target particle, and is given by the Compton equation:

$$\Delta \lambda = \frac{h}{m_0 \cdot c_r} \cdot (1 - \cos \theta), \tag{23}$$

where
$m_0$ = rest mass of the matter particle in its own frame, and
$c_r$ = magnitude of light velocity at a distance r from the center of the ECSF frame, that is, at the location of the experimental set up.

The constant portion in the above equation is called the Compton wavelength for the matter particle, and is expressed by:

$$\lambda_C = \frac{h}{m_0 \cdot c_r} = \frac{h \cdot c_r}{E_0}, \tag{24}$$

where
$E_0$ = rest mass-energy of the matter particle in its own frame where it is at rest,

If a photon-electron scattering experiment is designed to repeat with accelerated electrons then the above equation can be written as**:**

$$\lambda_r = \frac{h \cdot c_r}{E_r}, \qquad (25)$$

where

$E_r$ = mass-energy of the moving particle in its natural c.s.,

Dividing equations (25) by (24), and substituting the transformation factor for energy from equation (15) while noting that gravitational effect is negligible here, leads to

$$\frac{\lambda_r}{\lambda_0} = \frac{E_0}{E_r} = \sqrt{1 - \beta_r^2}, \qquad (26)$$

The wavelength $\lambda_r$ of the accelerated matter particle in the above experiment will give indirectly the order-of-magnitude measure of the diameter of the matter particle, and thus confirming the phenomenon of length contraction. The above equation proves that the transformation factor for length contraction is the same as that for energy.

### 3.3. Discussion

*3.3.1. Relativity of Simultaneity*

Einstein described [15] a thought experiment, in which a light signal is emitted from the center of a moving glass-walled room, which is observed by outside and inside observers. The outside observer infers that one of the walls is trying to escape from the light signal, and hence, the escaping wall will be met by the signal a little later than the approaching wall. Abandoning the viewpoint on simultaneity of classical physics, he introduced the relativistic viewpoint that two events which are simultaneous in one c.s., may not be simultaneous in another c.s.

The above thought experiment describes the concept of the relativity of simultaneity using the Einstein synchronization procedure — that is, using transmission of electromagnetic signals based on the principle of the constancy of the speed of light c — as laid down in SRT, and as applicable for inertial frames. This relativistic concept of simultaneity is an acceptable concept in RRT, but it uses the "remodeled Einstein synchronization procedure" that uses transmission of electromagnetic signals based on the principle of the natural velocity of light $c_r$. However, for cases (where c to $c_r$ variation is insignificant) like GPS application at its present accuracy level, for reasons explained earlier, the Einstein synchronization procedure using the constant speed of light can still be used.

It may be mentioned here that though the Einstein synchronization procedure is applicable for inertial frames, its use has limitations in case of rotating frames, due to the Sagnac effect. In this connection, Ashby mentions [19] that in a rotating reference frame, the Sagnac effect prevents a network of self-consistently synchronized clocks from being established by such processes.

In connection with the Sagnac effect, Ashby further mentions [19] that as a consequence of the Sagnac effect, observers in the rotating ECEF frame using Einstein synchronization will not agree that clocks in the ECI frame are synchronized, due to the relative motion. He also remarks [19] that observers in the rotating frame cannot in fact even globally synchronize their own clocks, due to the rotation.

Thus, it can be seen that even though GRT superseded SRT and is meant for tackling gravitational problems in rotating frames like GPS applications, the Einstein synchronization procedure cannot be used for synchronization of clocks in a rotating frame. Such paradoxical situation is really due to the fact that GRT has adopted the concept of "relativity of all frames". On the other hand, RRT avoids such paradox as it uses the concept of "nature's preferred frame".

*3.3.2. Length Contraction in the direction of the motion and a paradox*

Einstein [15] stated that it followed from the Lorentz transformation that a moving stick would contract only in the direction of the motion and the contraction would increase if the speed increases, and that there would be no contraction in the direction perpendicular to the motion.
This direction-dependant nature of length contraction, according to the present relativity theory, however, leads to a paradox as explained below.
Einstein stated [15] later that relativistic effects were caused by all kinds of energy, and that this conclusion, being quite general in character, was an important achievement of the theory of relativity and fitted all facts upon which it has been tested.
Thus, according to the present theory, in connection with relativistic effects due to translational kinetic energy, it may be said that nature's criterion for the direction of the length contraction is the direction of motion. Consequently, the question arises as to what is nature's criterion for the direction of the length contraction, for the cases**,** where this relativistic effect is caused by all other kinds of energy (viz., thermal energy, rotational kinetic energy, electromagnetic energy, etc.), than the translational kinetic energy?

According to RRT, all relativistic effects are caused by variation in energy level due to any kind of energy, and the contraction effect occurs along all directions. Thus, the paradox created by the present relativity theory on this issue does not arise in case of RRT.
It is noteworthy to mention here that even though the topic of length contraction of measuring rods or scales are so widely discussed, and are bound to occur according to any theory of relativity, the relativistic length contraction is so minute that it is not measurable by direct experiments for verification in any humanly conceivable experimental setup. This is primarily the reason why no direct experimental verification of length contraction has been reported so far till this centennial year of SRT. Detectable effects of length contraction can only occur at such energy levels at which the measuring scale can no longer exist but only its subatomic particles can. And, at those energy levels, one cannot talk of its length, breadth or height, but can attempt to devise a method to directly or indirectly measure its radius or diameter. Thus, according to RRT, the relativistic contraction occurs along all directions, whose indirect order-of-magnitude measure can be obtained in experiments involving the particle accelerators, as mentioned at subsections 3.2.2 to 3.2.4 above.

*3.3.3. Relative velocities and Limiting velocity for any material body*

Einstein stated [15] that from the point of view of his relativity theory the velocity of light formed the upper limit of velocities for all material bodies, and that the simple mechanical law of adding and subtracting velocities was no longer valid or, more precisely, was only approximately valid for small velocities, but not for those near the velocity of light.  He also mentioned [15] that the number expressing the velocity of light appeared explicitly in the Lorentz transformation, and played the role of a limiting case, like the infinite velocity in classical mechanics.

According to RRT, natural velocity of a material body or particle can be defined as a velocity with respect to its natural c.s., that is a consequence of nature's conservation laws,  and that is useful for precise calculation of similar other natural parameters, viz., kinematical time dilation, relativistic mass, relativistic length, etc. Relative velocity of a material body or particle with respect to any c.s. other than its natural c.s., is a parameter, that will not conform to the conservation laws; however, if it is to be utilized for any special or imprecise use, it should be done with proper caution and understanding of its limitations. Of course, the term 'natural velocity' should truly be called the 'natural relative velocity', because in fact it corresponds to the relative velocity with respect to the natural c.s. However, for simplification and for distinguishing it from the very commonly used term 'relative velocity' that has a long association with the concept of the "relativity of all c.s.", we prefer to use the term 'natural velocity'.

RRT also agrees to the view that the simple mechanical law of adding and subtracting velocities is neither valid nor useful for any precise calculation. This natural velocity can be calculated either by appropriately using equations (13) or (15) above, or by numerically simulating the journey of a material body or particle, in its relevant natural c.s.

According to RRT, the limiting velocity of a material body is not equal to 299,792,458 m/sec but equal to $c_r$ at the location corresponding to the body's radial coordinate r from the center of the gravitational natural c.s., whose origin coincides with the center of the gravitating mass, within whose sphere of influence the body exists at that particular instant.

### 3.4. General discussion

Lorentz transformation was developed from electromagnetism, and derived based on rectilinear motion in an inertial c.s., from a thought experiment involving propagation of the electromagnetic wave itself.
However, the role of Lorentz transformation in relativity theory is over-emphasized, because of the following reasons**:**
- it provided the earliest method for the theoretical calculation of the transformation factor for the relativistic effects like the kinematical time dilation, length contraction, etc.
- it had to play the important role for eliminating the contradiction between the first and the second postulate of SRT, and

- it had to play another important role for the second postulate of SRT by transforming the parameters from one inertial c.s. to another.

Originally developed for electromagnetism, Lorentz transformation has limited use for other areas of physics, say, for relativistic gravitation; and, when used for explaining any particular phenomena of electromagnetic field, it should be applied appropriately and from the point of view of natural c.s.

As has been shown using RRT in subsection 4.1.1 below, such transformation of the parameters is not useful or necessary for relativistic gravitation, where the calculations need to be done with respect to only the natural c.s., which is limited in extent by the sphere of influence.

Einstein [15] elaborated two thought experiments to emphasize on the relationship between the "relativity of all frames", and the relativistic effects of time dilation and of length contraction. For the first, he remarked [15] that the same result could certainly be found if the clock moved relative to an observer at rest in the upper c.s. instead of an observer at rest in the lower c.s., and that the laws of nature must be the same in both c.s. moving relative to each other. For the second, he remarked [15] that one could well imagine that not only did the moving clock change its rhythm, but also that a moving stick changed its length, so long as the laws of the changes were the same for all inertial c.s.

According to RRT, the relativistic effects occur (as explained below) for only one of the two c.s., which is at a higher energy level with respect to its natural c.s., according to the energy conservation calculations done following the methodology of nature**:**

- The macroscopic clock experiments including GPS have clearly confirmed the RRT contention that time dilation occurs only for the moving clocks, which are at a higher energy level than the clocks that are at rest in ECSF frame.
- SRT experiments have amply proven the phenomenon of relativistic mass-increase. RRT based simulations for orbits of photon and celestial bodies have amply proven this effect. But the SRT experiments as well as the RRT based simulations clearly indicate that the relativistic mass-increase occurs only for a body or a particle, which is at a higher energy level because of its motion with respect to its natural c.s.
- As regards length contraction, no direct experiment has proven it so far. However, an indirect order-of-magnitude measure of this phenomenon has been obtained for relativistic particles as mentioned at subsections 3.2.2 to 3.2.4 above. But even this indirect measure is obtainable for particles that are at a higher energy level because of their motion with respect to their natural c.s.

Summarizing for all the three relativistic effects, both experiments and simulations prove the RRT contention, and do not prove the SRT contention that these occur 'symmetrically for both c.s. moving relative to each other'.

According to RRT, the direction-dependant nature of length contraction was a consequence of the derivation method of Lorentz transformation based on rectilinear motion in an inertial c.s.

According to RRT, all relativistic effects are caused by variation in energy level due to any kind of energy, and the contraction effect occurs along all directions. The relativistic contraction of a spinning disc is governed by its rotational kinetic energy level with respect to its natural c.s., and occur along all directions. Thus, RRT has a Euclidean geometry for space, which has been applied and proven to be true from numerical simulation results using our photon and matter models [7, 8] as explained earlier.

### 3.5. Comparison of Concepts of the Present and the Proposed Relativity Theories

Summarizing the points discussed above, it becomes clear that some of the concepts that have gone into SRT and GRT need to be reviewed. RRT takes a comprehensive look at all the relativity experiments and proposes revised concepts that lead to a consistent relativity theory, in the table below.

| **SRT or GRT** | **RRT** |
| --- | --- |
| The relativistic effects occur for a body or a particle**:**<br>• due to its relative velocity with respect to any other body; the effect is symmetrical for both bodies moving relative to each other.<br>• due to gravitational effect of any other body; | The relativistic effects occur for a body or a particle**:**<br>• due to its natural velocity with respect to its natural c.s., which is a consequence of its higher energy level; the effect is caused for only the subject body.<br>• due to gravitational effect of the central body present in its natural c.s., as well as other perturbing bodies, which significantly affect its energy level; |
| Relative velocity of a material body or particle with respect to any c.s. is useful for relativistic calculations. | Natural velocity of a material body or particle with respect to only its natural c.s. is useful for relativistic calculations. |
| "Relativity of all frames" or "no preferred frame" is a basic premise of the present relativity theory. | "Nature's preferred frame" or natural c.s. is a basic premise of the proposed relativity theory (RRT). |
| The laws of nature must be invariant with respect to the appropriate transformation law, that is, the Lorentz transformation. | The conservation laws of energy, and of linear and angular momentum, and the laws of gravitation can and should be suitably applied appropriately following nature's methodology with respect to the relevant natural c.s. In other words, these laws need not be invariant with respect to any transformation law. |
| Length contraction of a body occurs only along the direction of motion, and no contraction occurs in the transverse direction. | Depending on energy level of a body with respect to only its natural c.s., contraction occurs along all directions irrespective of its direction of motion. |
| Limiting velocity of a material body is equal to the velocity of light. | Limiting velocity of a material body is equal to |

|  | the light velocity $c_r$ at its location in its gravitational natural c.s., where the body exists at that particular instant. |
|---|---|
| The simple mechanical law of adding and subtracting velocities is no longer valid or, more precisely, is only approximately valid for small velocities, but not for those near the velocity of light. | The velocity (that is, natural velocity) can be obtained either from equations (13) or (15), or from precise numerical simulation of a material body's trajectory in its natural c.s. |

## 4.   Coordinate Frames and Philosophical Principles of General Relativity

Regarding Einstein's approach to GRT as compared to SRT, Bernstein [23] mentioned that the early Einstein papers seemed to be rooted in a sort of clairvoyant view of the meaning of physical phenomena, and that there was an overriding sense of being close to the physical phenomena even when they were being described in new and apparently revolutionary terms; whereas, in the leap that led him to GRT, the connection with the phenomena was exceedingly indirect, and he was no longer guided by experiments (which came several years after the theory had been published), but by philosophical or epistemological principles.

This section discusses such principles that had gone into GRT, and presents the RRT views on the same, based on the enriching experience of almost a century of relativistic experimentation, and of almost half a century of space-age experimentation.

### 4.1.   Discussion

#### 4.1.1.   *The Galilean relativity principle*

Based on the Galilean relativity principle, Einstein mentioned [15] that a c.s. moved uniformly, relative to a "good" c.s., that is, one in which the laws of mechanics were valid. Perhaps, the situation was not ripe during that period for defining a c.s. as a "good" or "bad" one, based on such an intuitive idea of comparing the motion of an 'ideal train moving uniformly along a straight line', with say, the case of a 'train turning a curve'.

While acknowledging the fact that Galileo made extremely significant contributions for laying the foundation of physics and particularly for the foundation of mechanics, it may be appreciated that the Galilean relativity principle was concerned only with the 'linear' or 'rectilinear' motion. The lack of development of laws related to rotational motion, caused the development of relativity principles or theories to focus mainly on the rectilinear motion. Starting from the Galilean era, this trend continued much beyond the Newtonian era. In fact, even during the Einsteinian era, this trend was so strong that SRT, the first relativity theory that met rigorous experimentation, was devoted only to rectilinear motion, and linked to inertial c.s. In SRT, this developmental legacy of relativity theory made relativistic effects appear as a consequence of rectilinear motion or as a manifestation of translational kinetic energy and led to over-emphasis on the Lorentz transformation law. Even after Einstein gave final shape to GRT after 'expelling the ghost of inertial c.s.', this legacy left a rather indirect but lasting impression in his GRT, by making its geometry non-Euclidean. In fact, even at that time, there existed only sketchy ideas about one of the two principal types of motion, namely, the principle of 'rotational' inertia and motion — in the sense, that no generalized law of rotational motions like that in RRT was formulated, although there exists an underlying symmetry among the various phenomena involving spinning and rotational motions, viz., centrifugal force, Coriolis force, various precessional motions of gyroscopes and spinning tops, etc., as shown while deriving the RRT law at subsection 2.4.6 above and at Appendix I.

In the space age, computational capability and accuracy in celestial mechanics, enable us to simulate numerically the motion of a solar system planet or the Moon, moving along their perturbed curvilinear orbits, and make precision check of the conservation of the linear and angular momentum simultaneously and appropriately with respect to their relevant natural c.s. (say, the SCSF frame for the planetary orbits, or the ECSF frame for the lunar orbit) at every epoch, as explained in more detail in section 2 above. This proves that simultaneously these two conservation laws related to the to both the principal types of motion, can be used appropriately for computations with respect to the relevant natural c.s. That these laws related to both 'linear' and 'rotational' mechanics are valid in natural c.s., is proven from the fact that nature itself applies these conservation laws. Any observable change in the position of the orbiting body is the result of the combined effect on the motion of the body due to these conservation laws of linear and angular momentum as well as the conservation law of energy.

Thus, according to RRT, the rotating c.s. is not considered a "bad" c.s. When a rotating frame happens to be a natural frame, say for example, any real phenomenon occur on a spinning disk, it is tackled in RRT by using its generalized law of rotational motions.

Applying these laws with respect to any other c.s. is not correct, since the magnitudes of the parameters belonging to the equations of these laws when transformed to any other c.s., would not conform to the conservation laws. However, the laws are valid in any other c.s. for any other orbiting or moving body for which the other c.s. happens to be the natural c.s. The following example from application of this principle in precision computation of celestial orbits will make it clear.

For a planetary orbit of the solar system the SCSF frame is the natural c.s., since only in this c.s. the conservation laws are obeyed at every epoch.. Thus, the ECSF frame is not a natural c.s. for a planetary orbit of the solar system. Whereas, the ECSF frame is the natural c.s., for the orbits of the Moon or the GPS satellites

Thus, application in precision computation of celestial orbits prove the correctness of the aforesaid RRT principle that "the general forms of the laws of (rectilinear and rotational) motion — that is, the conservation laws of linear and angular momentum — are valid in all natural frames (that is, for both the SCSF and ECSF frames, as in case of the previous examples)", but these laws must be applied appropriately with respect to the relevant natural c.s.

If the ultimate check for the conservation laws is done with respect to any other c.s., using the transformed magnitudes of the parameters belonging to the equations of these laws, it will lead to non-compliance due to improper application of the conservation laws; however, this non-compliance will prove that observation of the phenomenon from this other c.s. is not a real but an apparent phenomenon. However, such magnitudes of the parameters (viz., the length, distance or time-intervals), that have been transformed to any other c.s. than the natural, may be useful for independent uses like determining the corrections to be applied for data obtained from apparent observations, checking of particle lifetimes, particle sizes, distances traversed by particles, etc.

A typical example is given below on how to obtain the positional coordinates at a particular instant of a body or particle moving under the gravitational influence of a celestial body. The positional coordinates at a particular instant of the Moon with respect to the ECSF frame, can be obtained precisely from the epochwise radius vector output generated by running a program that uses the RRT matter model [8]. Simpler method to transform time and length parameters is to use the energy-level transformation factors as discussed in sections 2 and 3.

Let's say an event (x, y, z, t) in a "natural c.s.", is being observed from a frame moving at velocity v along the direction of x-axis, relative to the former c.s. According to RRT, the corresponding (x', y', z', t') in this moving frame can be expressed in the form of a typical set of kinematical equations of SRT, by the following explicit functions

$$x' = [x - v.t] \bigg/ \left[\left(1 - \frac{G.M_C}{c_r^2 .r}\right) \cdot \sqrt{1 - \beta_r^2}\right]. \tag{27}$$

$$y' = y \bigg/ \left(1 - \frac{G.M_C}{c_r^2 .r}\right). \tag{28}$$

$$z' = z \bigg/ \left(1 - \frac{G.M_C}{c_r^2 .r}\right). \tag{29}$$

$$t' = \left[t - x.\left(v/c_r^2\right)\right] \bigg/ \left[\left(1 - \frac{G.M_C}{c_r^2 .r}\right) \cdot \sqrt{1 - \beta_r^2}\right]. \tag{30}$$

$$r = \sqrt{x^2 + y^2 + z^2}. \tag{31}$$

where

$M_C$ = mass of the central body of the relevant natural c.s.,

$\beta_r$ = velocity ratio of the moving or orbiting body at a radial coordinate r,
given by $\beta_r = v / c_r$,

r = radial distance from the origin of the relevant natural c.s., which coincides with the center-of-mass of the central body, and other notations being the same as given earlier.

### *4.1.2. Inertial forces and Mach's principle*

Rejecting the Newtonian concept in connection with the inertial forces, Einstein adopted the Machian concept in his GRT almost eight decades ago. He remarked [24] in this connection that the acceleration, which figured in Newton's equations of motion, was unintelligible if one started with the concept of relative motion, and that this fact compelled Newton to invent a physical space in relation to which acceleration was supposed to exist. He also asserted that this ad-hoc introduction of the concept of absolute space though logically correct, was unsatisfactory. He [24] also agreed with Mach's attempt to alter the mechanical equations in such a way that the inertia of bodies was traced back to relative motion on their part not as against absolute space, but as against the totality of other ponderable bodies.

Sciama [25] elaborates on Mach's views as follows**:**

> "Mach's criticism of Newton's laws of motion are more detailed than Berkeley's, but as regards centrifugal force his standpoint is the same. In 1872, Mach wrote**:**
>> 'For me <u>only relative motions exist</u>. … When a body rotates relatively to the fixed stars, centrifugal forces are produced, when it rotates relatively to some different body and not relative to the fixed stars, no centrifugal forces are produced. I have no objection to just calling the first rotation so long as it be remembered that nothing is meant except relative rotation with respect to the fixed stars.'
> In one point he is more explicit than Berkeley, when he says**:**
>> 'Obviously it does not matter if we think of the earth as turning round on its axis, or at rest while the fixed stars revolve round it. Geometrically these are exactly the same case of a relative rotation of the earth and the fixed stars with respect to one another. But if we think of the earth at rest and the fixed stars revolving round it, there is no flattening of the earth, no Foucalt's experiment, and so on — at least according to our usual conception of the law of inertia. Now one can <u>solve the difficulty in two ways</u>. Either all motion is absolute, or our law of inertia is wrongly expressed. I prefer the second way. The law of inertia must be so conceived that exactly the same thing results from the second supposition as from the first. By this it will be evident that in its expression, regard must be paid to the masses of the universe.' …
> According to Mach, then, inertial frames are those which are unaccelerated relative to the 'fixed stars' — that is, relative to some suitably defined average of all the matter of the universe. Moreover, matter has inertia only because there is other matter in the universe. Following Einstein, we shall call these statements Mach's principle."

As can be seen from the above, the basis of Mach's views spring from the concept of "relativity of all frames", to abandon which no strong reason could be found by Einstein before the space age developments.
RRT could find a third way to 'solve the difficulty' mentioned by Mach using its 'generalized vector law of rotational motions'. This law that generalizes not only all inertial forces but also all inertial torques, helped RRT to avoid both the Newtonian concept of 'absolute space' and the Machian concept of inertial forces as mentioned in connection with the centrifugal forces. It may be mentioned that according to RRT, the accelerations on a body exist not with respect to the 'absolute space', but with respect to its natural c.s., with which this body continuously does energy balance, as explained at subsection 2.4.2.
Since, Einstein incorporated the Machian concept in his GRT almost eight decades ago, no precision computation of planetary or lunar orbits has been reported in the relevant literature till date, based on Mach's view on inertial forces. Whereas, as explained at subsection 2.4.6 and above, we have successfully used for such computations the equations for inertial forces based on this RRT law.

*4.1.3. Freely Falling Frames and Locally Inertial Frames*

In an effort to develop the arguments for GRT, Einstein [15] had brought in the concepts of 'freely falling frames' and 'locally inertial frames' so as to at least indicate a c.s., in which all the physical laws are valid. For this, he has described a thought or idealized experiment and had stated [15] that a c.s. rigidly connected with the freely falling lift differed from the inertial c.s. (as represented in classical physics) in only one respect that it was limited in space and time, and thus was only a "pocket edition" of a real inertial c.s. He added that this local character of the c.s. is quite essential and the dimensions of the lift must be limited <u>so</u> that the equality of acceleration of all bodies falling freely with the lift, relative to the outside observer might be <u>assumed</u>. He also added that thus the c.s. could take on an inertial character for the inside observer, and one could at least indicate a c.s., in which all the physical laws were valid. He argued that one could imagine another similar c.s., that is, another lift moving uniformly relative to the one falling freely, then both these c.s. would be <u>locally inertial</u>, with all laws exactly the same in both, and the transition from one to the other would be given by the Lorentz transformation. Einstein noted that while the outside observer noticed the motion of the lift and of all bodies in the lift, he would find them in agreement with Newton's gravitational law, and thus for him, the motion would not be uniform, but accelerated due to the action of the gravitational field of the earth. However, he added that a generation of physicists born and brought up in the lift would believe themselves in possession of an inertial system, and would refer all laws of nature to their lift, stating with justification that the laws take on a specially simple form in their c.s., and hence it would be natural for them to <u>assume</u> that their lift was at rest and that their c.s. was the inertial one.

In a bid to develop the typical GRT arguments that extends equivalence principle (EP) for a broader equivalence between the laws of physics in different accelerated reference frames, Einstein [15] emphasized that it would be impossible to settle the differences between the outside and the inside observers, as each of them could claim the right to refer all events to his c.s., and concluded that both descriptions of events could he made equally consistent, but for such a description one must take into account gravitation, building a "bridge" which effected a transition from one c.s. to the other.
Thus, Einstein [15] concluded that the gravitational field existed for the outside observer; it did not for the inside observer, and that accelerated motion of the lift in the gravitational field existed for the outside observer, whereas, absence of motion and gravitational field for the inside observer.

Judging from present day precision level, it can be said that some imprecise arguments have been used here for describing the physical phenomena in two different c.s. The claim that 'both descriptions of events could he made equally consistent' is not tenable, because of the following reasons.
Applying the restriction that 'the dimensions of the lift must be limited so that the equality of acceleration of all bodies relative to the outside observer may be assumed', it has been assumed that the c.s. takes on an inertial character for the inside observer. Such approximation to support the assumption of 'equality of acceleration' may be valid for some special cases with only a few digits calculation accuracy, but very imprecise for 16 to 18 digit calculations that are done for celestial orbits in space age.

A quantitative view of the above qualitative statement can be taken from the following example. Let us calculate the maximum height of a lift, falling freely under the influence of the gravitational field of the earth, in which the velocity of two objects dropped for a free fall simultaneously from a state of rest, is being compared. One of the objects located just below the roof, and the other is located little above the floor of the lift. Both the objects are falling freely along with the lift. To avoid making a significant error in precision computations, the difference in velocity of these two objects should be smaller than $10^{-18}$ meters per second. For a location of the lift close to the earth surface, let us assume that the height of the lift is of the order of about three meters. For an elevation difference of two meters between the two bodies at the instant they are dropped, the difference in their velocities will be higher than $10^{-18}$ meters per second, only after about $1.6 \cdot 10^{-13}$ second. Thus, the errors incurred by approximations of this order over repeated steps would add up to significant level. If such order of error is inherent in the relativistic gravitational model used for computation, then the error incurred will affect the accuracy of the least-square adjusted constants, and ultimately will act as an impediment for further accuracy improvement in the orbit computations.

The above-mentioned GRT approach that describe a physical phenomenon observed from two different c.s., to effect a transition from one c.s. to the other, has a drawback that the inference drawn by the inside observer, which is nothing but an apparent observation, is considered to be a real phenomena of nature.
In this connection, one may recall a real life historical situation of the theory on geocentric Ptolemaic view of motions in the solar system based on apparent astronomical observations. This theory described the orbital motions of the sun, planets and the Moon around the earth in complex orbits. This theory, though quite complex, was accepted for almost fifteen centuries till the beginning of Copernican era, when it could not quantitatively account for an increasing number of observations. The modern practice in astronomy, as we all know, is to apply proper corrections to all apparent optical observations, before using such data to fit any orbital calculation or theory.
This and similar other experiences have taught us that if in a scientific theory apparent observations are given the status of true or real observations, it may result in a complicated theory, whose inconsistency will be revealed at a later date.
Hence, RRT avoids using any apparent observation as a real one, and also avoids making any approximation in the theoretical and mathematical models that may affect the significant digits in precision computation, as it may pose a limitation for the model. If at all for any particular case, an approximation is helpful, it should be done carefully after checking that it is not affecting the significant digits.

*4.1.4. Equivalence of Accelerated Motion and Motion in a Gravitational Field*

In a bid to apply the extended EP arguments for this case, Einstein [15] described the case of a lift that was being pulled with a constant force along the direction of motion, with respect to an inertial c.s. He explained that in the outside observer's view, his c.s. was an inertial one, the lift moved with a constant acceleration, the observers inside were in absolute motion, and for them the laws of mechanics were invalid; and, in the inside observer's view, his c.s. was not an inertial one, since his lift was in a gravitational field.

It may be mentioned here that in the accelerated lift, the acceleration is precisely same everywhere, and the vector direction of acceleration at different points of an accelerated c.s., are precisely parallel. Whereas, in a gravitational field, the magnitude of the acceleration varies from point to point, and the vector directions of gravitational acceleration at different points are radial all the way from the origin of the natural c.s. The degree of these variations is so much that such approximation is not acceptable for the precision demanded by celestial orbit computations in space age.

Einstein [15] described another thought experiment in the same lift by considering that a beam of light was sent in a horizontal direction, to check whether it met the wall at a point exactly opposite to that at which it entered. He argued that if the inside observer reasoned correctly and took into account the bending of light rays in a gravitational field, then his results would be <u>exactly</u> the same as those of an outside observer.

It may be mentioned here that precise analysis will show that the two observation results will not be exactly the same. The observed trajectory of a photon in the light-ray by the inside observer under a real gravitational field is supposed to be an arc of an hyperbolic orbit for photon that is supposed to be obtained from precision computation using RRT photon model [7]; but for this case of an unreal gravitational field, the simulation will fail and show non-compliance with the conservation laws due to improper application of the conservation laws; this non-compliance will also prove that observation of the phenomenon by the inside observer is not a real but an apparent phenomenon.

Whereas, the observed apparent bending by the outside observer (if numerically simulated for doing an accurate calculation) will lead to a bending angle depending on the initial velocity (at the instant the photon crosses the first wall), uniform acceleration and the distance traversed by the photon in the lift.

Einstein [15] argued further that one could thus eliminate "absolute" from such examples of non-uniform motion, by substituting with a gravitational field, and the ghosts of absolute motion and inertial c.s. could then be expelled from physics and a new relativistic physics would be built. In a bid to develop the extended EP arguments, Einstein [15] added that such idealized experiments showed how the problem of the GRT was closely connected with that of gravitation.

In the analysis of both these thought experiments, again both of the previous approaches — that is, treating an apparent observation as a real one, and making approximations in the theoretical model — have been utilized as evident from the above discussions.

RRT avoids such practices as it leads to a complicated and inconsistent theory as mentioned earlier.

*4.1.5. Geometry for Space-time and Equivalence of Centrifugal and Gravitational Accelerations*

Einstein [26] argued that it was necessary to frame a theory whose equations kept their form in the case of non-linear transformations of the coordinates. He felt that bringing in non-linear transformations (that is, generalizing the Lorentz transformations to include accelerations) as needed by the principle of equivalence, was inevitably destructive to the simple physical interpretation of the co-ordinates — in the sense that it could no longer be required that differentials of co-ordinates should signify direct results of measurement with ideal scales or clocks. This fact bothered him much, as it took a long time for him to see what co-ordinates in general really meant in physics. He could not find the way out of this dilemma [26] till 1912.

Einstein [15] describes the idealized experiment with the spinning disk, based on the concept of Lorentz contraction which owes its origin to Lorentz transformation law that was derived for inertial c.s., and that is associated with a paradox (as explained at subsection 3.3.2) created by the concept of direction-dependent length contraction. Einstein stated that the outside observer would find the ratio of the circumferences equal to that of the radii, since Euclidean geometry would be valid in his inertial c.s. From the point of view of classical physics and also the SRT, the inside observer's c.s. would be a forbidden one. Einstein applied the extended EP arguments, and added that one must treat the observer on the disk and the observer outside with equal seriousness. Einstein concluded [15] that if one wished to reject absolute motion and to keep up the idea of the GRT, then physics must all be built on the basis of a geometry more general than the Euclidean.

Based on the above thought experiment, Einstein [15] had explained the GRT standpoint through an imaginary conversation between the outside observer, having belief in classical physics, and the inside observer, with belief in GRT and in Mach's principle, who would not have anything to do with absolute motion. According to the latter, his c.s. was just as good as that of the classical physicist, and he noticed only other's rotation relative to his disk, and hence no one could forbid him to relate all motions to his disk.

The reply of the general relativist was as though he had certainly noticed the two facts mentioned by the classical physicist; he held a <u>strange</u> gravitational field (which acted in a radial but outward direction) acting on his disk, responsible for them both. The gravitational field, being directed toward the outside of the disk, deformed the relativist's rigid rods and changed the rhythm of the relativist's clocks. The gravitational field, non-Euclidean geometry, clocks with different rhythms are, for the relativist, all closely connected. Accepting any c.s., the relativist must at the same time <u>assume</u> the existence of an appropriate gravitational field with its influence upon rigid rods and clocks.

During the time GRT was formulated, such approximation might have been acceptable. But today at the level of space age precision, it will not be possible to devise a gravitational field that will be both accurate and precisely equivalent, since the gravitational and centrifugal acceleration terms in the ODE's will lead to different results from numerical integration. Also, if such numerical integration is done using the RRT, the first one consisting of the gravitational term will prove that it is an apparent phenomenon, when checks for conservation laws are carried out; whereas, numerical integration of the second one will prove that it is a real phenomenon, when checks for conservation laws are carried out (as mentioned earlier, numerical integration of the ODE's involving inertial acceleration terms are regularly done for both the photon and matter models of RRT).

As explained above, RRT replaced the concept of "relativity of all frames" with that of "nature's preferred frame". Thus, any real phenomenon on a spinning disk or a rotating frame, will be tackled in RRT by using its generalized law of rotational motions.

*4.1.6.   A Concluding Remark on GRT*

In a bid to sum up the basis for his GRT using arguments that extends EP for a broader equivalence between the laws of physics in different accelerated reference frames, Einstein [15] stated that these idealized experiments (as discussed in the previous three subsections) indicated only the general character of the new relativistic physics, that they showed one that the fundamental problem was that of gravitation, and that they also showed one that GRT led to further generalization of time and space concepts. Einstein [15] concluded that thus the solution of the gravitational problem in the GRT must differ from the Newtonian gravitation and, hence, the laws of gravitation must be formulated for all possible c.s., just as all laws of nature.

At the beginning of the twentieth century Einstein adopted the concept of "relativity of all c.s." based on the legacy of Galilean relativity principle, as there was no strong reason to disfavor this concept. The fact that non-Euclidean space was its compelling consequence is clear from his comments [15], that 'there exists no way of escape from this consequence (*of non-Euclidean geometry*) if all c.s. are permissible'.
Whereas, in the space age with the advantage of improvements in technological and computational capabilities, we found how nature is applying the conservation laws in case of a real phenomenon of nature, and applied RRT following the methodology of nature. This approach helps us to identify not only the "nature's preferred frame", but also the real phenomenon from an apparent phenomenon.

*4.1.7.   Evolution of Einstein's Concepts to Curved Space-time*

Bernstein [23] stated that Einstein's 1911 paper [27] was an attempt to set the equality of gravitational and inertial mass into a more general framework, though it was not a fully successful attempt, as Einstein had still not abandoned Newton's theory of gravitation, and as he was adjoining some new extra principles in the old Newtonian theory in an amalgam that did not stick together. Bernstein [23] further mentioned that some of the concepts Einstein introduced then have survived, and the paper was interesting for what it revealed about the evolution of Einstein's thought, and that one could sense Einstein's struggle in a way that one could not in the 1916 paper, which has an austere and finished perfection.

Before abandoning Newton's theory of gravitation Einstein had tried for considerable time to formulate the GRT model by adjoining some new extra principles to the former. But, it was not successful because he had no strong reason to abandon the concept of "relativity of all c.s.", as is clear from our earlier discussions.

Einstein had the dilemma born out of the direction-dependent nature of Lorentz transformations. But, the role of Lorentz transformation in relativity theory is over-emphasized because of the reasons mentioned at subsection 3.4 above. The phenomenal experimental success of SRT during the decade he worked on the formulation of GRT, also did not allow him to cast a doubt on this transformation law or any of its features, although no answer was available to the question**:** what is nature's criterion for the direction of the length contraction, for those cases where the relativistic effect is caused by all kinds of energy other than the translational kinetic energy.

Thus, until 1912, Einstein himself was much bothered by this dilemma, but he could neither get a strong reason to abandon Lorentz transformations, nor could find its alternative that can help to calculate the transformation factors for the relativistic effects as in SRT.

The remodeling effort that we started had the advantage of almost eight decades of experiments on the various relativistic effects. We sought after the more fundamental principles that may be the cause for the relativistic effects, and adopted the experimentally proven principle that energy level (due to all forms of energy) is the more fundamental underlying cause behind these effects. From this principle we calculated the transformation factors for the relativistic effects, and proved them consistently (as explained in sections 2 and 3 above) using the experimental or observational data from various relativistic experiments. Thus, we proved that even for SRT experiments, the relativistic transformation factors could be calculated without using the Lorentz transformation.

*4.1.8. RRT and Euclidean Geometry for Space*

We took a hint from Einstein's struggle as mentioned in the previous subsection, and followed the nature's methodology. We gave importance not to the laws of mechanics, but to the conservation laws of momentum, which are the more general forms of the laws of mechanics, and which are more fundamental than the laws of mechanics in the sense that they are valid in both the macroscopic and microscopic realms.
We modified the Newtonian equations for gravitation by using the well-founded relativistic principles, and developed a photon model and a matter model, in a consistent manner.
These models were used to numerically simulate respectively the orbits for photon as well as celestial bodies in the solar system including the Moon. The computation results of the photon and matter models [7, 8] compare well with the recent experimental results as explained earlier.
The programs for precision computation of planetary and lunar orbits as well as of photon orbit ensure that the conservation laws of energy, linear and angular momentum are obeyed during all stages of the calculations. This check for conservation laws helps us to identify the natural c.s., because only when we run our program with respect to the "nature's preferred frame" the simulations converge and the conservation laws are obeyed, and consequently we can truly be said to follow the methodology of nature.

Since, unlike the case of GRT, the concepts of "relativity of all c.s." and Lorentz transformation or its paradoxical feature, viz., direction-dependent nature of length contraction did not form a part of RRT, there was no compulsion or logic for RRT to give up the concept of Euclidean space.
Even though, the concept of Euclidean space in RRT, is a consequence of its fundamental principles, this concept was proven true from successful computation results, using our photon and matter models [7, 8] as explained earlier.

It may also be mentioned that the experimental effects of geodetic bending or precession, ascribed to the concept of space-time curvature according to the GRT, are accurately accounted for as the corresponding relativistic effects in RRT, as presented in our papers [7, 8].

Since, the Euclidean space of the RRT model, may lead to some questions from the point of view of curved space-time of the GRT, these are being addressed to, in the following few paragraphs of this subsection.

The first question is whether the RRT model goes back to the old principle of "action at a distance"? The answer is that in the post-space-age scenario, RRT model certainly does not go back to this old principle or any of its other forms. We consider gravity as a propagating force of nature in flat space-time whose propagation speed should be derived from observational evidences. We are working on this aspect of the RRT model. For the work presented in this and our earlier papers [7, 8], we considered the error incurred due to the propagation speed of gravity, to be of such order that it is accommodated within the uncertainty levels of present-day observations and computations.

The second question is**:** whether the theories of black holes can be set up using the RRT model? It can be answered by stating that since black holes are a consequence of the gravitational collapse of large stellar masses, their formation and existence is not dependent on the gravitational model. Of course, the theory of black holes for the RRT model needs to be set up differently as compared to the GRT model. We have started our work on this, and, in due course of time will publish a paper on the same.

The third question is**:** whether the theories of cosmology can be set up using the RRT model? It can be answered in the affirmative by stating that for obvious reasons, the cosmological theory, using the RRT model, needs to be set up differently as compared to the GRT model. In this connection it may be mentioned that some cosmologists (e.g., Magueijo, etc.) are already working on cosmological models using the VSL concept, which has helped them to make better progress.

### 4.2. General Discussion

While formulating GRT, Einstein
- abandoned the inertial frame of Galilean relativity;
- concluded that there 'seems to be no reason' to abandon the concept of "relativity of all frames";
- continued to over-emphasize the role of Lorentz transformation, though he found that the consequence of bringing in non-linear transformations for generalizing the Lorentz transformations to include accelerations, would 'be fatal to simple interpretation of co-ordinates' and, would lead to the modification of the geometry of space-time;
- incorporated Mach's concept that the inertia of bodies is traced back to relative motion on their part as against the totality of other ponderable bodies. (Mach's view also had a strong influence from the concept of "relativity of all frames".)

In connection with inertia, Mach [25] was right in stating, "Either all motion is absolute, or our law of inertia is wrongly expressed. I prefer the second way.", but his law of inertia was not right.  Based on his stated view — "For me only relative motions exist. … When a body rotates relatively to the fixed stars, centrifugal forces are produced, when it rotates relatively to some different body and not relative to the fixed stars, no centrifugal forces are produced." — no precision computation of planetary or lunar orbits has been reported in the relevant literature till date. Hence, in RRT, the Machian concept of inertial forces was replaced with a generalized vector law of rotational motions.

From the detailed discussions above, it can be concluded that adoption of the concept of "relativity of all c.s." in GRT, led to the consequence of non-Euclidean space. RRT replaced this concept with that of "nature's preferred frame", since only on such a frame nature does conserve energy, linear and angular momentum, and also because the following compelling reasons supported this view**:**
- results of the macroscopic clock experiments and experience of GPS applications (since the concerned relativists had no option but to adopt the ECSF frame),
- experience on precision ephemeris generation has taught us that today one has to accept the existence of the astronomical constants (e.g., the planetary masses, etc.) of nature as a concomitant of only one appropriate preferred frame and relevant orbit or orbits, linked to them,
- importance of "nature's preferred frame" can also be understood from the energy-balance point of view based on equation (6) above, as explained at subsection 2.4.2, and
- experience of ephemeris generation by astronomers has shown that it is useful to employ GRT equations in a "preferred frame" for planetary and lunar orbits.

Hence, RRT abandoned the concept of "relativity of all frames", that was central to both Einstein's theories and Mach's principle, and adopted the concept of "nature's preferred frame", and this helped the authors to satisfy Einstein's unfulfilled desire to retain 'the simple interpretation of co-ordinates'.

## 5. Conclusion

RRT, supported by strong reasons, adopted the concept of "nature's preferred frame" in lieu of the concept of "relativity of all frames".

The fact that variation in energy level is the deeper underlying cause for relativistic effects, was adopted as one of the fundamental principles of RRT. This principle helped the authors to formulate RRT as a more consistent theory avoiding the Lorentz transformation, that has paradoxical direction-dependent nature.

Among other fundamental principles adopted in RRT are the conservation laws of energy, linear and angular momentum, and the well-proven relativistic energy equation from Einstein.

With these few basic principles, it was possible to formulate RRT with Euclidean space, which could then be used consistently and successfully for numerical simulation of the results of "well-established" tests of GRT at their current accuracy levels, and also for the precise calculation of relativistic effects observed in case of GPS applications, the accurate macroscopic clock experiments and other tests of SRT, as presented in this paper and our earlier papers [7, 8].

We proved in our photon model [7] the principle mentioned by Einstein [15] that 'a beam of light will bend in a gravitational field exactly as a (*material*) body would if thrown horizontally with a velocity equal to, that of light'.

Thus, the RRT photon model enables one to numerically simulate the magnitude of the variable speed of light, and to confirm Einstein's assertion [21] that light rays could curve only when the velocity of propagation of light varies with position.

Thus, it can be concluded that retaining the essential essence of Einstein's theories and enriching them with the knowledge gained from experimental and the space-age developments, RRT has been formulated to avoid their inadequacies.

## Appendix I

### Generalized vector law of Spinning and Rotational Motions

The following table presents the equations that relate the external forces or torques with the rate of change in the direction of linear and angular momentum vectors, and finally presents the generalized vector law that reveals the high degree of symmetry among all forms of rotational motions.

| | Linear Momentum | Angular Momentum | | Remarks |
|---|---|---|---|---|
| | | **Spin** | **Orbital** | |
| Equations for momentum vector | $\mathbf{P} = m \cdot \mathbf{v}$ | $\mathbf{L}_s = I_s \cdot \dot{\boldsymbol{\alpha}}_s$ | $\mathbf{L}_o = \mathbf{r} \times \mathbf{P}$ | |
| where | $m$ = relativistic mass, $\mathbf{v}$ = rectilinear velocity vector | $I_s$ = moment of inertia about the spin or symmetry axis passing through the body's center of mass; and, $\dot{\boldsymbol{\alpha}}_s$ = spin angular velocity vector. | $\mathbf{r}$ = radius vector, and, $\mathbf{P}$ = linear momentum vector | |
| Equations for external force or torque given by the Vector law of momentum | $\mathbf{F} = \dfrac{d\mathbf{P}}{dt}$ | $\boldsymbol{\tau}^s = \dfrac{d\mathbf{L}_s}{dt}$ | $\boldsymbol{\tau}^o = \dfrac{d\mathbf{L}_o}{dt}$ | |
| Equations for the external force or torque, split into longitudinal and transverse components | $\mathbf{F} = \mathbf{F}_{\|} + \mathbf{F}_{\perp}$ | $\boldsymbol{\tau}^s = \boldsymbol{\tau}^s_{\|} + \boldsymbol{\tau}^s_{\perp}$ | $\boldsymbol{\tau}^o = \boldsymbol{\tau}^o_{\|} + \boldsymbol{\tau}^o_{\perp}$ | Longitudinal and transverse direction means parallel and perpendicular to the momentum vector direction, respectively (Note 1). |
| Equations for the external force or torque, when the transverse component is zero and only the longitudinal component exists | $F_{\|} = \dfrac{dP}{dt}$ | $\tau^s_{\|} = \dfrac{dL_s}{dt}$ | $\tau^o_{\|} = \dfrac{dL_o}{dt}$ | These are the conservation laws of momentum vector magnitude. |
| Equations for the external force or torque, when the longitudinal component is zero and only | $\mathbf{F}_{\perp} = \dfrac{d\mathbf{P}}{dt}$ | $\boldsymbol{\tau}^s_{\perp} = \dfrac{d\mathbf{L}_s}{dt}$ | $\boldsymbol{\tau}^o_{\perp} = \dfrac{d\mathbf{L}_o}{dt}$ | These are vector forms of the conservation laws of momentum |
| the trans-verse component exists | | | | vector direction |
| Names of inertial forces and torques for each case | Centrifugal force; Coriolis force | Gyroscopic resistance; and Precessional resistance of a spinning Top | Inertial torque that opposes turning of orbital plane | |
| This is the finally derived vector form of the conservation law of momentum vector direction | $F_\perp = \dot{\boldsymbol{\theta}} \times P$ | $\boldsymbol{\tau}_\perp^s = \dot{\boldsymbol{\theta}} \times \mathbf{L}_s$ | $\boldsymbol{\tau}_\perp^o = \dot{\boldsymbol{\theta}} \times \mathbf{L}_o$ | Derivations of a few sample cases have been shown below. (Note 2) |
| where | $\dot{\boldsymbol{\theta}}$ = vector representing the rate of change of momentum vector direction with respect to the space-fixed frame. | | | |
| **Example:** | The orbital motion of a body in a two-body circular orbit. | The precessional motion of a spinning top (Figure 1) or a typical gyroscope mounted on gimbals on a cantilever frame (Figure 2) | The precessional motion of a planetary orbit in solar system | |
| The external force or torque that tends to change the momentum vector direction | $F_\perp$ is caused by the centripetal force due to gravitational attraction | $\boldsymbol{\tau}_\perp^s$ is caused by the weight of the spinning top or the heavy spinning gyroscope (Note 3) | $\boldsymbol{\tau}_\perp^o$ is caused by the relevant component of the perturbative force | |
| **Symmetry of the law of conservation of momentum vector direction, in vector form** | | | | |
| Conservation law of momentum vector direction, in vector form | $F_\perp = \dot{\boldsymbol{\theta}} \times P$ | $\boldsymbol{\tau}_\perp^s = \dot{\boldsymbol{\theta}} \times \mathbf{L}_s$ | $\boldsymbol{\tau}_\perp^o = \dot{\boldsymbol{\theta}} \times \mathbf{L}_o$ | It is a fundamental law in the sense that it is valid even in the relativistic domain. |
| | | $\boldsymbol{\tau}_\perp^s$ and $\mathbf{L}_s$ | $\boldsymbol{\tau}_\perp^o$ and $\mathbf{L}_o$ | Hereafter, super- and sub- scripts have been dropped from the LHS symbols. |
| Conservation law of momentum vector direction, in vector form | $F_\perp = \dot{\boldsymbol{\theta}} \times P$ | $\boldsymbol{\tau}_\perp = \dot{\boldsymbol{\theta}} \times \mathbf{L}$ | | |
| Generalized expression of the conservation law of momentum vector direction expressed as a vector cross product  *External torque or force* = (*Rate of change of Momentum Vector Direction*) × *Momentum* | | | | |

**Note 1.** In most cases both the longitudinal and transverse force or torque components will be present. But, there can be two limiting cases where only one of two components will be present, the magnitude of the other being zero. Thus, in a case where the longitudinal component is zero, it is easy to understand the transverse component, because as shown in the subsequent few rows, in such a situation the conservation law of momentum relates to one of the two split parts of the force or torque, namely the transverse component, which we refer to here as the 'vector law of conservation of momentum vector direction', and finally generalize as the 'generalized law of spinning and rotational motions'. The other limiting case helps one to understand the longitudinal component that relates to the 'vector law of conservation of momentum vector magnitude', as shown in the next row.

**Note 2.** The derivation of the equations for forces and torques given below, use a right-handed spherical ($r$, $\theta$, $\Phi$) coordinate system and the mutually perpendicular unit vectors $\hat{i}$, $\hat{j}$, and $\hat{k}$, corresponding respectively to the $r$-, $\theta$-, and $\Phi$- directions.

**Centrifugal force:**
$$\mathbf{F}_\perp = \frac{d\mathbf{P}}{dt} = m_r r \dot{\theta} \frac{d\hat{j}}{dt} = m_r r \dot{\theta} \left(\dot{\boldsymbol{\theta}} \times \hat{j}\right) = \dot{\boldsymbol{\theta}} \times \mathbf{P}$$

**Torque that resists change of the vector direction of Orbital Angular Momentum:**
$$\boldsymbol{\tau}^o_\perp = \frac{d\mathbf{L}_o}{dt} = \frac{d(\mathbf{r} \times \mathbf{P})}{dt} = m_r r^2 \dot{\theta} \frac{d(\hat{i} \times \hat{j})}{dt} = m_r r^2 \dot{\theta} \frac{d\hat{k}}{dt} = m_r r^2 \dot{\theta} \left(\dot{\boldsymbol{\theta}} \times \hat{k}\right) = \dot{\boldsymbol{\theta}} \times \mathbf{L}_o$$

where

$m_r$ = relativistic mass of the moving or orbiting body (say, a planet) at a radial coordinate r in its natural c.s., as given by equation (4) at subsection 3.3 above.

The torque given in the following equations uses a right-handed coordinate system and the mutually perpendicular unit vectors $\hat{i}$, $\hat{j}$, and $\hat{k}$, corresponding respectively to radial, tangential, and vertical directions.

**Gyroscopic resistance:**
$$\boldsymbol{\tau}^s_\perp = \mathbf{a} \times m\mathbf{g} = -mga\left(\hat{i} \times \hat{k}\right) = mga\,\hat{j}, \text{ and}$$

$$\boldsymbol{\tau}^s_\perp = \frac{d\mathbf{L}_s}{dt} = L_s \frac{d\hat{i}}{dt} = L_s \left(\dot{\boldsymbol{\theta}} \times \hat{i}\right) = \dot{\boldsymbol{\theta}} \times \mathbf{L}_s$$

**Note 3.**  The external torque that tends to change the momentum vector direction, is caused by the weight of the spinning top or the heavy spinning gyroscope, respectively, as can be seen in the following two equations, that have been derived using conventional methods in scalar form and presented in the references mentioned, and which agree with relevant equations presented in the table above in vector form.

|  | **Spinning Top** | **Spinning Gyroscope** |
|---|---|---|
| Equations for the rate of change of momentum vector direction | $\dot{\theta} = \dfrac{\tau_{\perp e}}{L_s} = \dfrac{mg \cdot r}{L_s}$ | $\dot{\theta} = \dfrac{\tau_{\perp e}}{L_s} = \dfrac{mg \cdot a}{L_s}$ |
| Reference | Resnick and Halliday [28] | Crabtree [29] and Sears [30] |

It may be mentioned here that it is possible to discern the vector law of conservation of momentum vector direction, from the vector diagrams (Figures 1b and 2b) for the spinning top and the spinning gyroscope. The same vector law can also be observed to exist in the vector diagrams for the other two cases, viz., for linear momentum and orbital angular momentum.

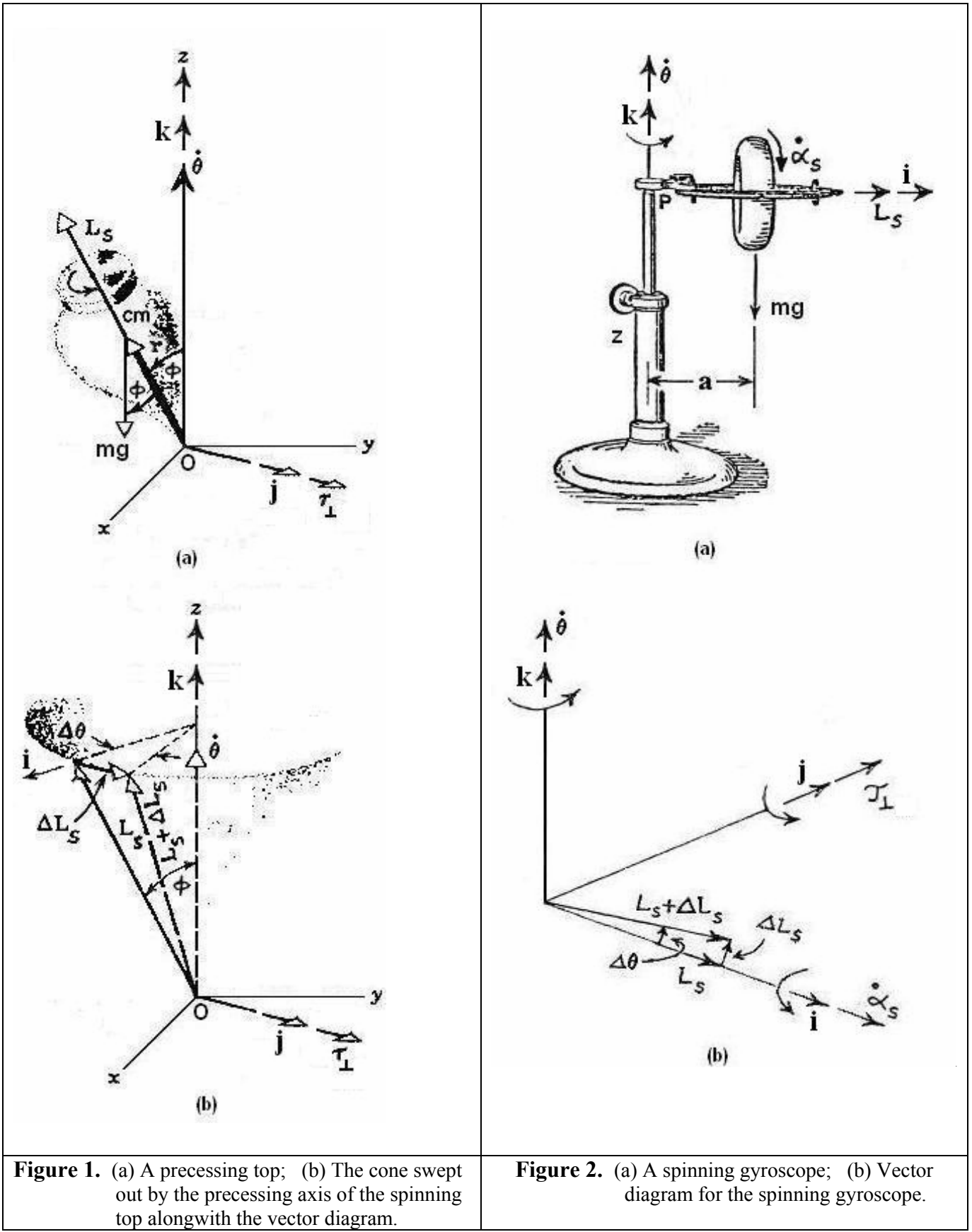

**Figure 1.** (a) A precessing top; (b) The cone swept out by the precessing axis of the spinning top alongwith the vector diagram.

**Figure 2.** (a) A spinning gyroscope; (b) Vector diagram for the spinning gyroscope.